\documentclass[11pt,twocolumn,tighten,resetfootnote]{aastex7}

\usepackage{apjfonts}
\usepackage{url}
\usepackage{hyperref}
\usepackage{natbib}
\usepackage{amsmath,amstext,amssymb}
\usepackage[caption=false]{subfig} % for subfloat
\usepackage{caption}
\DeclareCaptionLabelFormat{moviefmt}{#1~#2\ (Movie)}
\usepackage{xcolor, fontawesome}
\usepackage{color}
\usepackage{enumitem}
\usepackage{rotating} % sidewaystable

\linenumbers % turn line numbering on or off

\newcommand{\kms}{\ensuremath{\rm km\,s^{-1}}}

\newcommand{\sint}{\sigma_{\mathrm{int}}}
\newcommand{\chired}{\chi^2_{\nu}}
\newcommand{\dBIC}{\Delta\mathrm{BIC}}
\newcommand{\um}{\,\mu\mathrm{m}}

\newcommand{\ipac}{Caltech/IPAC, Pasadena, CA 91125, USA}
\newcommand{\carnegie}{Carnegie Science Observatories, Pasadena, CA 91101, USA}
\newcommand{\caltech}{Department of Astronomy, Caltech, Pasadena, CA 91125, USA}
\newcommand{\standrews}{School of Physics and Astronomy, University of St Andrews, St Andrews, Scotland KY16 9SS, UK}

\newcommand{\stsci}{Space Telescope Science Institute, Baltimore, MD 21218, USA}
\newcommand{\cfa}{Center for Astrophysics $\vert$ Harvard \& Smithsonian, Cambridge, MA 02138, USA}
\newcommand{\mitkavli}{Kavli Institute for Astrophysics and Space Research, MIT, Cambridge, MA 02139, USA}

\newcommand{\komaba}{Komaba Institute for Science, The University of Tokyo, 3-8-1 Komaba, Meguro, Tokyo 153-8902, Japan}
\newcommand{\abc}{Astrobiology Center, 2-21-1 Osawa, Mitaka, Tokyo 181-8588, Japan}
\newcommand{\naoj}{National Astronomical Observatory of Japan, 2-21-1 Osawa, Mitaka, Tokyo 181-8588, Japan}
\newcommand{\sokendai}{Astronomical Science Program, Graduate University for Advanced Studies, SOKENDAI, 2-21-1 Osawa, Mitaka, Tokyo 181-8588, Japan}
\newcommand{\okayama}{Okayama Observatory, Kyoto University, 3037-5 Honjo, Kamogatacho, Asakuchi, Okayama 719-0232, Japan}
\newcommand{\utokyograd}{Department of Multi-Disciplinary Sciences, Graduate School of Arts and Sciences, The University of Tokyo, 3-8-1 Komaba, Meguro, Tokyo 153-8902, Japan}
\newcommand{\iac}{Instituto de Astrof\'isica de Canarias (IAC), 38205 La Laguna, Tenerife, Spain}
\newcommand{\ull}{Departamento de Astrof\'isica, Universidad de La Laguna (ULL), 38206 La Laguna, Tenerife, Spain}
\newcommand{\berkeley}{Astronomy Department, University of California, Berkeley, CA 94720, USA}
\newcommand{\ticthreeveq}{89.7} % 2piRstar/Prot
\newcommand{\lptwelveveq}{24.2} % 2piRstar/Prot
\newcommand{\dgcvnveq}{136} % 2piRstar/Prot
\newcommand{\fourstarobsratio}{0.80}           % observed depth ratio NB2.09/TESS
\newcommand{\fourstarobsratiounc}{0.03}        % observed depth ratio unc
\newcommand{\fourstarpredratio}{0.64}          % predicted ratio, with sigma_int
\newcommand{\fourstarpredratiounc}{0.05}       % predicted ratio unc, with sigma_int
\newcommand{\fourstarpredrationosig}{0.63}     % predicted ratio, no sigma_int
\newcommand{\fourstarpredrationosigunc}{0.01}  % predicted ratio unc, no sigma_int
\newcommand{\nunderpredictsigma}{2.9}          % NIR under-prediction [sigma], with sigma_int (fiducial)
\newcommand{\nunderpredictsigmanosig}{5.9}     % NIR under-prediction [sigma], no sigma_int
\newcommand{\brokenbetaone}{0.79}              % broken power law beta1 (blueward), with sigma_int [fiducial]
\newcommand{\brokenbetaoneunc}{0.16}           % broken beta1 unc, with sigma_int
\newcommand{\brokenbetatwo}{-0.09}             % broken power law beta2 (redward), with sigma_int
\newcommand{\brokenbetatwounc}{0.39}           % broken beta2 unc, with sigma_int
\newcommand{\brokenlambdabreak}{1.18}          % break wavelength [um], with sigma_int
\newcommand{\brokenlambdabreakunc}{0.27}       % break wavelength unc [um], with sigma_int
\newcommand{\brokenlambdabreaknosig}{1.22}     % break wavelength [um], no sigma_int
\newcommand{\brokenlambdabreaknosigunc}{0.31}  % break wavelength unc [um], no sigma_int
\newcommand{\singlechirednosig}{8.39}          % single power law chi2_nu vs meas err, no sigma_int
\newcommand{\brokenchirednosig}{7.25}          % broken power law chi2_nu vs meas err, no sigma_int
\newcommand{\dbicsinglebroken}{6.7}            % dBIC(single) - dBIC(broken), with sigma_int
\newcommand{\dbicsinglebrokennosig}{51.2}      % dBIC(single) - dBIC(broken), no sigma_int
\begin{document}

\title{Conspicuous Gas, Cryptic Dust: Spectroscopy and Chromaticity of Complex Periodic Variables}

\correspondingauthor{Luke G. Bouma}

\received{2026~August~31}
\revised{---}
\accepted{---}
\shorttitle{Conspicuous Gas, Cryptic Dust} 

\shortauthors{Bouma et al.}

\author[orcid=0000-0002-0514-5538,sname='Bouma']{Luke~G.~Bouma}
\affiliation{\ipac}
\affiliation{\carnegie}
\email{lbouma@ipac.caltech.edu}

\author[orcid=0000-0002-4128-7867,sname='P\'erez Paolino']{Facundo~P\'erez~Paolino}
\affiliation{\caltech}
\email{fperezpa@caltech.edu}
 
\author[sname='Hillenbrand']{Lynne~A.~Hillenbrand}
\affiliation{\caltech}
\email{lah@astro.caltech.edu}
 
% --- Tierras ---
\author[orcid=0000-0003-1361-985X,sname='Garc\'ia-Mej\'ia']{Juliana~Garc\'ia-Mej\'ia}
\altaffiliation{51 Pegasi b Fellow, MIT Pappalardo Physics Fellow}
\affiliation{\mitkavli}
\affiliation{\cfa}
\email{jgarciam@mit.edu}
 
% --- KeplerCam ---
\author[orcid=0000-0001-6637-5401,sname='Bieryla']{Allyson~Bieryla}
\affiliation{\cfa}
\email{abieryla@cfa.harvard.edu}

\author[orcid=0000-0003-2171-5083,sname='Tamburo']{Patrick~Tamburo}
\affiliation{\cfa}
\email{patrick.tamburo@cfa.harvard.edu}
 
\author[orcid=0000-0002-9003-484X,sname='Charbonneau']{David~Charbonneau}
\affiliation{\cfa}
\email{dcharbonneau@cfa.harvard.edu}
 
% --- MuSCAT1 (Japanese side) ---
\author[orcid=0000-0002-4881-3620,sname='Livingston']{John~H.~Livingston}
\affiliation{\abc}
\affiliation{\naoj}
\affiliation{\sokendai}
\email{john.livingston@nao.ac.jp}
 
\author[orcid=0000-0002-6480-3799,sname='Isogai']{Keisuke~Isogai}
\affiliation{\okayama}
\affiliation{\utokyograd}
\email{isogai@kusastro.kyoto-u.ac.jp}
 
\author[orcid=0000-0001-8511-2981,sname='Narita']{Norio~Narita}
\affiliation{\komaba}
\affiliation{\abc}
\affiliation{\iac}
\email{narita@g.ecc.u-tokyo.ac.jp}
 
\author[orcid=0000-0002-4909-5763,sname='Fukui']{Akihiko~Fukui}
\affiliation{\komaba}
\affiliation{\iac}
\email{afukui@g.ecc.u-tokyo.ac.jp}
 
% --- MuSCAT2 (Spanish side) ---
\author[orcid=0000-0003-0987-1593,sname='Pall\'e']{Enric~Pall\'e}
\affiliation{\iac}
\affiliation{\ull}
\email{epalle@iac.es}
 
\author[orcid=0000-0001-9087-1245,sname='Murgas']{Felipe~Murgas}
\affiliation{\iac}
\affiliation{\ull}
\email{fmurgas@iac.es}

% --- HIRES (helped observing 20240216) ---
\author[orcid=0000-0002-9540-853X,sname='Carvalho']{Adolfo~S.~Carvalho}
\affiliation{\caltech}
\affiliation{\cfa}
\email{adolfo.carvalho@cfa.harvard.edu}

% --- HIRES (reductions) ---
\author[0000-0002-0531-1073,sname='Isaacson']{Howard~Isaacson}
\affiliation{\berkeley}
\email{hisaacson@berkeley.edu}

%%%%%%%%%%

\author[orcid=0000-0002-5547-3775,sname='Spake']{Jessica~J.~Spake}
\affiliation{\carnegie}
\email{jspake@carnegiescience.edu}

\author[orcid=0000-0001-6381-515X,sname='Rebull']{Luisa~M.~Rebull}
\affiliation{\ipac}
\email{rebull@ipac.caltech.edu}

\author[orcid=0000-0002-1466-5236,sname='Jardine']{Moira~M.~Jardine}
\affiliation{\standrews}
\email{mmj@st-andrews.ac.uk}
 
\author[orcid=0000-0002-5741-3047,sname='Ciardi']{David~R.~Ciardi}
\affiliation{\ipac}
\email{ciardi@ipac.caltech.edu}

\author[orcid=0000-0003-3305-6281,sname='Valenti']{Jeff~A.~Valenti}
\affiliation{\stsci}
\email{valenti@stsci.edu}

%%%%%%%%%%%%%%%%%%%%%%%%%%%%%%%%%%%%%%%%%%%%%%%%%%%%%%%%%%%%%%%%%%%%%%%%%%%%%%%
%%%%%%%%%%%%%%%%%%%%%%%%%%%%%%%%%%%%%%%%%%%%%%%%%%%%%%%%%%%%%%%%%%%%%%%%%%%%%%%

\begin{abstract}
  Complex periodic variables (CPVs) are young low-mass stars whose
  light curves show periodic dips indicative of transiting corotating
  material.  The origin and composition of this material are unclear.
  Here we present new optical and near-infrared spectroscopy and
  photometry of four CPVs from Magellan, Keck, Hale, MuSCAT1, MuSCAT2,
  Tierras, KeplerCam, and TESS.   The spectra imply that CPVs host
  magnetically bound circumstellar plasma clumps, on the basis of
  sinusoidal-in-time Balmer emission out of transit, and Balmer
  dimming during transit.  Yet large night-to-night changes in
  circumstellar hydrogen emissivity occur without clear changes in
  light curve morphology, suggesting that the sharp flux dips are
  caused not by circumstellar plasma but by dust.  Optical
  chromaticities ($\delta \propto \lambda^{-\beta}$,
  $\beta$=\brokenbetaone$\pm$\brokenbetaoneunc) support this, but the
  power law breaks in the near-infrared, where a single power law
  under-predicts the depths observed at 2.1\,$\mu$m.  We therefore
  favor dips caused by dusty plasma clumps with opaque cores and
  optically thin halos, though we cannot rule out models in which the
  dust properties vary per-star or per-epoch.  Observations at
  $\lambda$>2\,$\mu$m and $\lambda$<$0.4$\,$\mu$m would test this
  interpretation and clarify the dust's origin.
\end{abstract}

\keywords{Circumstellar matter (241), Stellar magnetic fields (1610),
Stellar rotation (1629), Periodic variable stars (1213), Weak-line T
Tauri stars (1795)}

%%%%%%%%%%%%%%%%%%%%%%%%%%%%%%%%%%%%%%%%%%%%%%%%%%%%%%%%%%%%%%%%%%%%%%%%%%%%%%%
%%%%%%%%%%%%%%%%%%%%%%%%%%%%%%%%%%%%%%%%%%%%%%%%%%%%%%%%%%%%%%%%%%%%%%%%%%%%%%%

\clearpage
\newpage
\section{Introduction}
\label{sec:intro}

Complex periodic variables (CPVs) present the kind of signal that
astronomers usually find easy to explain: large amplitude, short period, and
repeatable.  Yet a decade after their discovery, the composition,
structure, and origin of the occulting material remain unknown.  CPVs
are young M dwarfs, typically with rotation periods shorter than two
days and, within current sensitivity limits, no disks
\citep{Stauffer2017,Bouma2024}.  The material trapped in their
magnetospheres transits at a few stellar radii: the region occupied by
the closest planets, and an environment that is otherwise hard to
measure.  While the impact of stellar photons on exoplanets is widely
appreciated \citep{France2016,Owen2019}, the particle environment can
be just as influential \citep{Catling2017}, and means for studying it
are scarce \citep{Loyd2025}.  CPVs may provide such a means.

\begin{figure}[!t]
  \centering
  \includegraphics[width=0.48\textwidth]{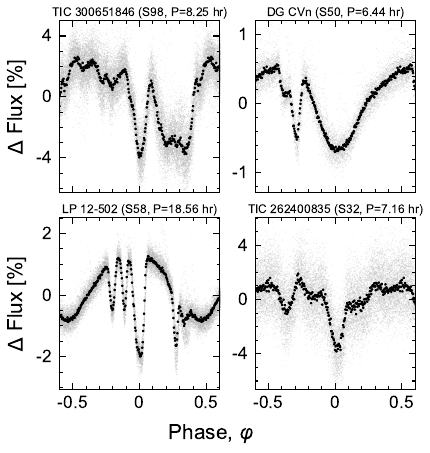}
  \vspace{-0.4cm}
  \caption{{\bf TESS light curves of the four targets.}  We acquired
  new spectra and multicolor photometry for TIC~300651846, DG~CVn
  (TIC~368129164), LP~12-502 (TIC~402980664), and TIC~262400835.
  Panels show one month of TESS data folded on each object's period,
  with 120-second cadence flux measurements in gray and their average
  in black.  For DG~CVn, we removed a best-fit sinusoid to the
  $P_2$=2.60\,hr secondary companion's rotation signal (3.1\,AU
  projected separation; \citealt{Beuzit2004}).  The time evolution of
  each target from 2018--2025 is shown in the associated
  \href{https://lgbouma.com/cgcd_2026}{{\bf online movies}}: while
  the periods remain fixed, the morphologies evolve over months to
  years.}
  \label{fig:lcmosaic}
\end{figure}

Figure~\ref{fig:lcmosaic} shows the main objects of interest in this
article; over one hundred analogous systems have been reported by K2
and TESS
\citep{Rebull2016,Stauffer2017,Rebull2018,Zhan2019,Rebull2020,Stauffer2021,Palumbo2022,Popinchalk2023,Bouma2024}.
CPVs are identified from their structured, periodic optical light
curves.  About 3\% of stars a few million years old show the behavior,
declining to $\approx$0.3\% by $\approx$110\,Myr \citep{Rebull2020}.
The leading interpretation is that clumps of charged circumstellar
material corotate with the star \citep{Stauffer2017}.  The physical
requirement is a ``centrifugal magnetosphere''
\citep[e.g.][]{Petit2013}: when the magnetospheric radius $R_{\rm m}$
exceeds the corotation radius $R_{\rm c}$, the effective potential
along closed field lines acquires local minima at radii beyond $R_{\rm
c}$, where the field enforces super-Keplerian motion and centrifugal
support confines material against infall \citep{Townsend2005}.  The
trapped material is thus {\it in} corotation with the star, while
residing {\it beyond} the corotation radius.

Spectroscopy offers a direct test of the corotating-material
hypothesis, and the necessary data have only recently become
available.  Early spectroscopic characterization established CPVs as
weak-lined T~Tauri stars: chromospherically active, with no evidence
for accretion \citep{Stauffer2017}.  The first time-series
spectroscopy of a CPV then showed excess H$\alpha$ emission that
varied sinusoidally with rotation phase, tracing cool
($\lesssim$$10^4$\,K) plasma clumps at opposing stellar longitudes
near 2--4\,$R_\star$, consistent with a pinched torus of
magnetically trapped plasma \citep{Bouma2025}.  Simulations show that
such gas can condense around rapidly rotating cool stars, with the
outermost material being periodically ejected \citep{Daley-Yates2024}.
Comparable H$\alpha$ emission
\citep{Donati2000,Townsend2005b,Dunstone2006,Skelly2008} and an
analogous dimming as the clumps transit the stellar disk
\citep{CollierCameron1989,CollierCameron1992,Cang2020} are seen across
a wide range of masses in rapidly rotating magnetic stars.  Those
stars, however, do not show complex photometric variability, and
before this work, TIC~141146667 was the only CPV with unambiguous
signatures of bound circumstellar plasma, leaving the question of
universality open.  Do all CPVs harbor bound plasma?  Which of the
observed morphological features, such as opposing clumps,
night-to-night variability, or Balmer dimming during transits,
generalize across the class?

While spectroscopy can reveal the gas, multicolor photometry provides
independent evidence for what causes the dips.  Ground-based campaigns
have established that CPV dips deepen toward the blue, with power-law
slopes $\beta$$\approx$0--1.5 for $\delta \propto \lambda^{-\beta}$
\citep{Onitsuka2017,Tanimoto2020,Gunther2022,Koen2023}.  This behavior
suggests extinction by sub-micron dust.  Three issues, however,
complicate the interpretation.  First, published dip depths rest on
heterogeneous detrending choices, which propagate into the inferred
chromaticity.  Second, if the occulting material is partially opaque,
the observed slope is diluted relative to the intrinsic opacity
law, an effect that has so far been ignored.  Third, the models
considered have rarely extended beyond single power laws.  The
exception is the pioneering study of PTFO~8-8695 by
\citet{Tanimoto2020}, who explored broken extinction laws, albeit
interpreting that system's CPV \citep{Bouma2020} as a planet
candidate.  These limitations matter because the chromaticity directly
links to the question of where the dust in these systems comes from.
Are these M dwarfs condensing out dust within their own
magnetospheres, or trapping the detritus of planet formation?
Neither conclusion can be drawn without first establishing what the
occulting material does to starlight.

In this work we present time-series spectroscopy and multicolor
photometry of four CPVs (Figure~\ref{fig:lcmosaic}): 18 nights of
ground-based spectroscopy and nine nights of ground-based photometry
acquired between 2020 and 2025, usually with simultaneous TESS
coverage.  High-resolution spectroscopy at sensitivity comparable to
the TIC~141146667 observations yields a detection of sinusoidal
circumstellar Balmer emission for a second CPV, TIC~300651846, and
upper limits for LP~12-502.  Lower-resolution spectra with
simultaneous photometry show that Balmer dimming during broadband flux
dips is ubiquitous, though often with timing offsets.  For the
chromaticity, we homogeneously analyze our new photometry together
with the archival literature, placing 51 dip depths across five stars
on a common scale, and we compare four extinction law models.  Under
the assumption that a universal extinction law applies across the
population, a broken power law between the optical and near-infrared
is preferred; absent that assumption, models allowing star-to-star or
epoch-to-epoch variability fit as well or better.

Section~\ref{sec:obs} describes the observations;
Section~\ref{sec:results} interprets the
spectroscopy and fits possible extinction laws; Section~\ref{sec:disc}
discusses the structure and possible origins of the occulting
material; Section~\ref{sec:conc} concludes.

\section{New Observations}
\label{sec:obs}

\begin{deluxetable*}{lcccccccc}
\tablewidth{0pt}
\tabletypesize{\small}
\tablecaption{Overview of Stellar Properties\label{tab:starparams}}
\tablehead{
  \colhead{Star} &
  \colhead{$P$} &
  \colhead{$T_{\rm eff}$} &
  \colhead{$R_\star$} &
  \colhead{$M_\star$} &
  \colhead{Age} &
  \colhead{$v_{\rm eq}$} &
  \colhead{$R_{\rm c}$} &
  \colhead{Multiple?} \\[-5pt]
  \colhead{} &
  \colhead{(hr)} &
  \colhead{(K)} &
  \colhead{($R_\odot$)} &
  \colhead{($M_\odot$)} &
  \colhead{(Myr)} &
  \colhead{(\kms)} &
  \colhead{($R_\star$)} &
  \colhead{}
}
\startdata
TIC\,300651846 & 8.25  & $3159^{+56}_{-84}$ & $0.61^{+0.04}_{-0.03}$ & $0.40^{+0.02}_{-0.03}$ & 23  & $89.7 \pm 5.2$  & $2.49 \pm 0.15$ & N \\
LP\,12-502     & 18.56 & $3090^{+31}_{-59}$ & $0.37^{+0.02}_{-0.01}$ & $0.22^{+0.01}_{-0.02}$ & 80  & $24.2 \pm 1.0$  & $5.78 \pm 0.27$ & N \\
DG\,CVn        & 6.44  & $3127^{+51}_{-42}$ & $0.72^{+0.02}_{-0.03}$ & $0.40^{+0.02}_{-0.06}$ & 150 & $135.8 \pm 4.7$ & $1.79 \pm 0.09$ & Y \\
TIC\,262400835 & 7.16  & $3074^{+64}_{-58}$ & $1.00^{+0.04}_{-0.05}$ & $0.30^{+0.03}_{-0.04}$ & 19  & $169.6 \pm 7.6$ & $1.26 \pm 0.07$ & Y \\
\enddata
\tablecomments{DG\,CVn is TIC\,368129164; LP\,12-502 is
  TIC\,402980664.  Following \citet{Bouma2024}, radii and temperatures
  are from SED fits and masses are from PARSEC isochrones
  \citep{Chen2014}.  This procedure may yield biased results for binaries.  The LP~12-502 age
  comes from co-moving, co-spatial pre-main sequence stars
  \citep{Tofflemire2021}; the remaining ages are from
  association memberships \citep{Kerr2021,Gagne2026} and have relative
  uncertainties $\approx$$\pm$15\%.  
  % 300651846 CPL8 Gagne26
  % DG CVn CVUL Gagne26
  % LP 12-502 comove
  % TIC 262400835 greater Taurus Kerr2021
  The equatorial velocity $v_{\rm eq}$$=$$2\pi R_\star/P$ and the
  Keplerian corotation radius $R_{\rm c}$$=$$(G M_\star
  P^2/4\pi^2)^{1/3}$ are derived quantities.  The multiplicity column
  is Yes/No: DG\,CVn has a companion at $\rho$=3.1\,AU 
  \citep{Beuzit2004};
  TIC\,262400835 is likely an unresolved pair within a hierarchical
  triple (see text).}
\end{deluxetable*}

\subsection{Selection Function}
\label{sec:selfn}

We selected targets from the known CPV census
weighing brightness, period, morphology, accessibility, and
binarity
\citep{Rebull2016,Stauffer2017,Rebull2018,Zhan2019,Rebull2020,Stauffer2021,Palumbo2022,Popinchalk2023,Bouma2024}.
Brighter stars enable higher S/N.
Shorter periods maximize the fraction of a cycle covered per night,
and hence the probability of catching a sharp dip, since morphologies
are unpredictable beyond a few weeks.
Sharp, large-amplitude dips are most easily distinguished from
spot-induced variability.
Proximity to the TESS continuous viewing zones maximizes the number of
schedulable windows for simultaneous photometry and spectroscopy.
Finally, we were agnostic to binarity, since no CPVs with close
($<$0.1\,AU) companions capable of affecting the circumstellar
environment are known \citep[{\it cf.}][]{Curiel2026}.

These considerations yielded five targets.  LP\,12-502, DG\,CVn,
TIC\,300651846, and TIC\,262400835 have new data presented here;
TIC\,141146667 was discussed in \citet{Bouma2025}.
Table~\ref{tab:starparams} lists the stellar properties, and
Tables~\ref{tab:obs_spec} and~\ref{tab:obs_phot} the observations.

This work is the first to highlight TIC~262400835
(Gaia~DR3~3234917110482839680) as a CPV.  It was found in 2020 through
CDIPS \citep{Bouma2019} but omitted from the \citet{Bouma2024}
catalog, whose automated
methods did not recover it.  Its short period and complex
morphology motivated TESS DDT028, during which we also acquired
MuSCAT1 and MuSCAT2 photometry.
\citet{Kerr2021} placed it in the $\approx$19\,Myr ``Greater
Taurus Subgroup 3A''.
The system is likely a triple.
A resolved companion, TIC 262400836 (Gaia DR3 3234920065420339712),
lies 10$\farcs$6 west
($\sim$1{,}800\,AU projected) with consistent parallax and
proper motion.  The companion is 0.3\,magnitudes bluer in BP-RP at
near-identical $G$, and so is likely the primary.  
TIC~262400835 itself is likely an unresolved
pair, based on its Gaia DR3 RUWE (1.5) and photometric excess.  A
single 21$''$ TESS pixel shows three periods: the complex signal at
$P_1$=7.16\,hr, and sinusoidal signals at $P_2$=4.31\,day and
$P_3$=12.70\,hr.  The MuSCAT data (Section~\ref{sec:muscat}) localize
the 7.16\,hr signal to TIC~262400835, which suffices for this work.
\citet{Heinze2018} reported the resolved companion (TIC~262400836) as a $P$=4.3\,day close
binary; we recover the period but see no eclipses in the TESS data.

\begin{deluxetable*}{llllcccc}
\tablewidth{0pt}
\tabletypesize{\small}
\tablecaption{Spectroscopic Observations\label{tab:obs_spec}}
\tablehead{
  \colhead{Star} &
  \colhead{Spectrograph} &
  \colhead{Photometer} &
  \colhead{Night (UT)} &
  \colhead{Gas Resolvable?} &
  \colhead{Gas Emission?} &
  \colhead{H$\alpha$ Dips?} &
  \colhead{Flare Var.?}
}
\startdata
LP\,12-502                            & DBSP  & Tierras & 2023 Nov 11 & $\times$     & $\times$            & $\checkmark$ & $\checkmark$ \\
LP\,12-502                            & DBSP  & Tierras & 2023 Nov 12 & $\times$     & $\times$            & $\checkmark$ & $\checkmark$ \\
LP\,12-502                            & DBSP  & $\times$ & 2023 Dec 07 & $\times$     & $\times$            & -- & -- \\
LP\,12-502                            & DBSP  & TESS    & 2024 Nov 04 & $\times$     & $\times$            & $\checkmark$ & $\checkmark$ \\
LP\,12-502                            & DBSP  & TESS    & 2024 Nov 05 & $\times$     & $\times$            & $\times$ & $\checkmark$ \\
LP\,12-502                            & DBSP  & TESS    & 2024 Nov 06 & $\times$     & $\times$            & $\times$ & $\checkmark$ \\
LP\,12-502                            & DBSP  & TESS    & 2024 Nov 07 & $\times$     & $\times$            & $\checkmark$ & $\checkmark$ \\
LP\,12-502                            & DBSP  & TESS    & 2024 Nov 08 & $\times$     & $\times$            & $\times$ & $\checkmark$ \\
LP\,12-502                            & HIRES & $\times$ & 2023 Nov 23 & $\checkmark$ & $\times$            & -- & -- \\
LP\,12-502                            & HIRES & TESS    & 2024 Dec 03 & $\checkmark$ & $\times$            & $\checkmark$ & $\checkmark$ \\
LP\,12-502                            & HIRES & TESS    & 2024 Dec 10 & $\checkmark$ & $\times$            & $\times$ & $\checkmark$ \\
LP\,12-502                            & TSPEC & $\times$ & 2024 Jan 27 & $\times$     & $\times$            & -- & -- \\
DG\,CVn                               & DBSP  & Tierras & 2024 Apr 28 & ?            & $\times$            & $\checkmark$ & $\times$     \\
TIC\,300651846                        & MIKE  & TESS    & 2025 Jan 02 & $\checkmark$ & $\checkmark$        & $\checkmark$ & $\times$     \\
TIC\,300651846                        & MagE  & TESS    & 2025 Jan 18 & $\checkmark$ & $\checkmark$        & $\checkmark$ & ?            \\
TIC\,300651846                        & MagE  & TESS    & 2025 Jan 19 & $\checkmark$ & $\checkmark$        & $\checkmark$ & $\times$     \\
TIC\,300651846                        & MagE  & TESS    & 2025 Jan 20 & $\checkmark$ & $\times$            & $\checkmark$ & $\times$     \\
TIC\,300651846                        & MagE  & TESS    & 2025 Jan 21 & $\checkmark$ & $\times$            & $\checkmark$ & $\times$     \\
TIC\,141146667$^{\dagger}$            & HIRES & TESS    & 2024 Feb 17 & $\checkmark$ & $\checkmark$        & $\checkmark$ & ?     \\
\enddata
\tablecomments{ The last four columns indicate (i) whether the
  spectrograph had sufficient resolution relative to the stellar
  equatorial velocity to resolve bound circumstellar gas in emission,
  (ii) whether such gas was seen, (iii) whether the H$\alpha$
  equivalent width time series showed dips synchronous with broadband
  flux dips, and (iv) whether the spectra and photometry showed
  synchronous evidence for flare-induced variability.
  $^{\dagger}$Data for TIC~141146667 were
  described in \citet{Bouma2025}.}
\end{deluxetable*}

\subsection{Magellan/MagE}
\label{sec:mage}

\begin{figure*}[!tp]
  \centering
  \includegraphics[width=0.99\textwidth]{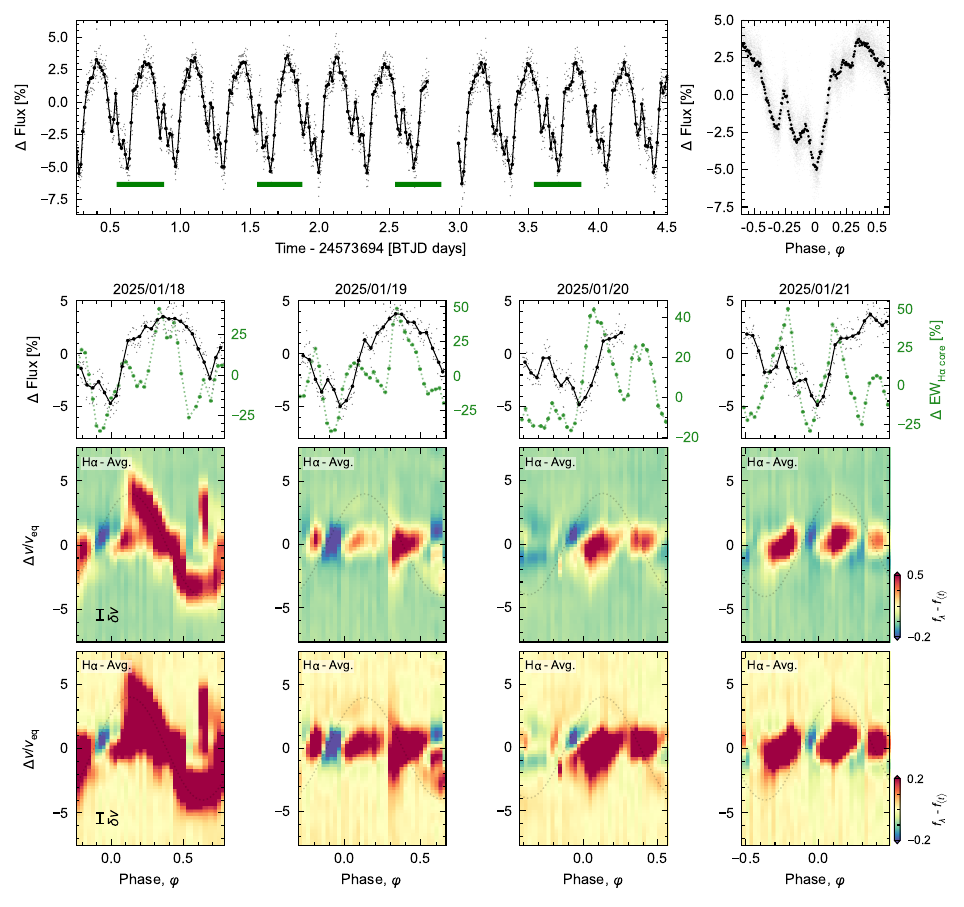}
  \vspace{-0.2cm}
  \caption{{\bf Circumstellar Balmer emission varies while the light
  curve does not.}  Data are from four successive nights on
  Magellan/MagE for TIC~300651846 ($P=8.25$\,hr).  The top-left panel
  shows TESS data (raw in gray; 20~minute time binning in black); the
  next row zooms in on times when MagE and TESS overlap, and shows
  the line-core H$\alpha$ equivalent width (at $|v/v_{\rm eq}|<1$)
  with green markers.  The lower two rows show time-series H$\alpha$
  profiles at two different color stretches, with the time-averaged
  mean profile subtracted.  The average TESS light curve from
  14~January to 11~February~2025 is in the top right.  Large
  amplitude sinusoidal-in-time Balmer emission is visible on
  18~January 2025.  Such emission is weakly visible from
  $\phi$=0.3-0.6 on 19~January and 20~January, and not present on
  21~January.  Despite the spectroscopic changes, the complex
  morphology in the TESS light curves remained stable.  The error bar
  in the lower left indicates the velocity resolution.  The faint
  model is a sinusoid with $K$=4\,$v_{\rm eq}$ and $\phi_0$=-0.12.
  }
  \label{fig:mage}
\end{figure*}

We observed TIC~300651846 ($V$=14.1) over four consecutive nights
spanning 18~January to 21~January~2025 using the Magellan Echellette
(MagE; \citealt{Marshall2008}) spectrograph on the Baade 6.5\,m
Magellan telescope.  We used the 1$''$$\times$10$''$ slit
($R$$\approx$4{,}100, $\delta v$$\approx$73\,\kms), with a 12 minute
cadence, and aligned the slit along the parallactic angle while taking
thorium-argon arcs for wavelength calibration once per hour.  The
seeing averaged 0$\farcs$8 over all four nights, rarely exceeding
1$''$.  We extracted 1-D spectra from the resulting 161 science-ready
echellograms using the CarPy data reduction package
\citep{Kelson2000,Kelson2003}, with S/N$\approx$160 per resolution
element in the continuum near H$\alpha$.  We timed these observations
to be simultaneous with TESS Sector 88.  Adopting the stellar radius
and period from \citet{Bouma2024}, TIC~300651846 has an equatorial
velocity $v_{\rm eq}$=\ticthreeveq\,\kms, implying near-critical
sampling of the line core.  Figure~\ref{fig:mage} shows the resulting
time-dynamic spectra in the vicinity of H$\alpha$, where the bottom
row subtracts the time-averaged spectrum from each night to highlight
variability---whether in the line core, or in the circumstellar
environment.

% S/N per pixel ~87, counts 7500.
% TESS s88 overlap: 3694.5 to 3697.8

\subsection{Magellan/MIKE}
\label{sec:mike}

\begin{figure}[!t]
  \centering
  \includegraphics[width=0.45\textwidth]{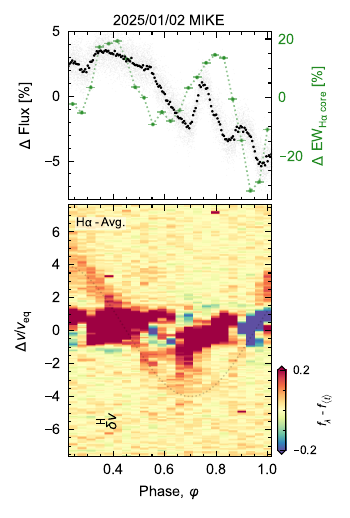}
  \vspace{-0.2cm}
  \caption{{\bf Magellan/MIKE H$\alpha$ time series of TIC~300651846
  on 2~January~2025.}
  These data have eight times higher resolution than the MagE data in
  Figure~\ref{fig:mage}; we bin by a factor of two in velocity for
  clarity, and again phase to $P=8.25$\,hr.  The circumstellar
  emission mirrors that seen on 19 and 20~January.
  Sinusoidal-in-time H$\alpha$ emits at $\approx$10\% of the
  continuum.  Absorption in the line core from $\phi$=0.9-1.0 occurs
  during the transit of the corresponding clump of plasma.  The
  ephemeris is the same as in Figure~\ref{fig:mage}; the average
  H$\alpha$ line profile (not shown) is similarly in emission with
  EW$=$$6.0 \pm 0.3$\,\AA.  The faint model is a sinusoid with
  $K$=4\,$v_{\rm eq}$ and $\phi_0$=-0.05, slightly offset from
  Figure~\ref{fig:mage}.
  }
  \label{fig:mike}
\end{figure}

We additionally observed TIC~300651846 on the night of 2~January~2025
using the Magellan Inamori Kyocera Echelle (MIKE;
\citealt{Bernstein2003}) spectrograph on the Clay 6.5\,m Magellan
telescope.  We used the 0.7$''$$\times$5$''$ slit, delivering
$R$$\approx$35{,}000 ($\delta v$$\approx$9\,\kms) in the red arm and
$R$$\approx$28{,}000 ($\delta v$$\approx$11\,\kms) in the blue, and
opted to read out unbinned in the spectral direction, and 2$\times$
binned in the spatial direction.  We bracketed every science
observation with a thorium-argon arc, since the main expected
advantage of this observation configuration relative to MagE was the
velocity resolution, which oversamples the line core.  The seeing over
the night averaged 0$\farcs$9.  We extracted 1-D spectra from the
resulting 23 science-ready echellograms using CarPy
\citep{Kelson2000,Kelson2003}, yielding S/N$\approx$54 per resolution
element in the H$\alpha$ continuum.  These observations were executed
simultaneously with TESS Sector 87.
Figure~\ref{fig:mike} shows the resulting time-dynamic spectra.

% 600 cts per px in red arm near Halpha continuum, 
% 5 pixels per resolution elemnt
% first observation DATE-OBS= '2025-01-03T00:46:19'        / (start)

\subsection{Hale/DBSP}
\label{sec:dbsp}

We used the Double Spectrograph (DBSP; \citealt{Oke1982}) on the
Palomar 200-inch telescope (Hale/P200) to observe LP~12-502 over seven
nights: 11~November to 12~November 2023, 7~December 2023, and
4~November to 8~November 2024.  We additionally observed DG~CVn
($V$=12.0), a known unresolved binary \citep{Osten2016} in which one
component is a CPV, over the night of 28~April~2024.  We used the
1$\farcs$5 slit, with the 1200 lines/mm grating, blazed at 5000\,\AA\
on the blue arm, and 1200 lines/mm, blazed at 7100\,\AA\ on the red
arm.  This delivered $R$$\approx$2{,}200 at H$\alpha$, corresponding
to $\delta$$v$$\approx$140\,\kms.  For LP~12-502, $v_{\rm
eq}$=\lptwelveveq\,\kms, yielding an experiment focused on equivalent
width measurement, rather than velocity resolution.  For DG~CVn
($v_{\rm eq}$=\dgcvnveq\,\kms) the velocity sampling near the line
core was near-critical.  For LP~12-502, we used 10 minute exposures,
yielding 442 spectra with 73.7 total hours of open-shutter time,
typically with S/N$\approx$200 per resolution element.  For DG~CVn, we
used five minute exposures, yielding 49 spectra over a 4.3 hour baseline.
We performed the extraction using \texttt{DBSP\_DRP}
\citep{dbsp_drp:joss}, which is an extension of the \texttt{PypeIt}
suite \citep{Prochaska2020}.  We then measured the equivalent widths
of the highest S/N Balmer lines---H$\alpha$ (n=3$\rightarrow$2),
H$\beta$ (n=4$\rightarrow$2), and H$\gamma$ (n=5$\rightarrow$2)---by
fitting the continuum-normalized spectra
with a one-component Gaussian with free amplitude, width, and mean.  We
estimated the uncertainties on each line width through a Monte Carlo
procedure that bootstrapped against the local scatter in the continuum. 

\begin{figure*}[p]
  \centering
  \includegraphics[width=0.98\textwidth]{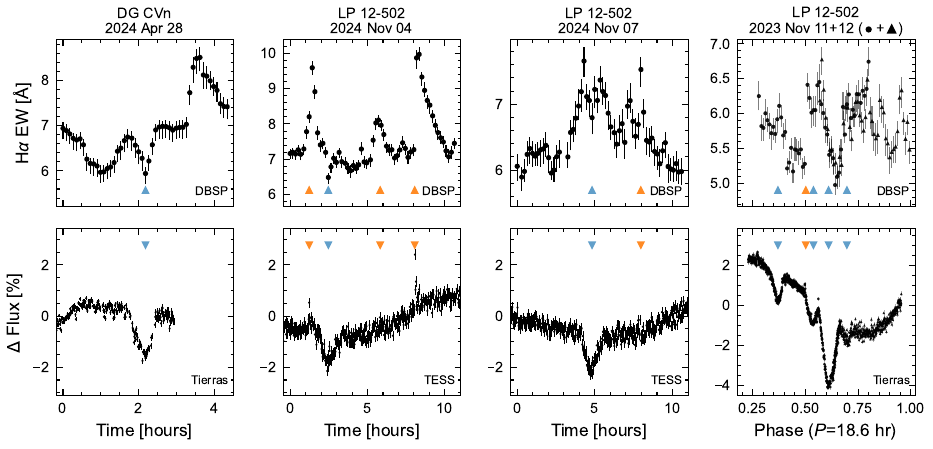}
  \vspace{-0.2cm}
  \caption{{\bf Balmer lines dim during dips and brighten during
  flares.}  Panels show Hale/DBSP ($R$$\approx$2{,}200) H$\alpha$
  equivalent widths (EWs) vs.~simultaneous Tierras and TESS
  photometry.
  Minute-timescale optical flares (orange arrows) are associated with
  hour-timescale Balmer EW flares, which dominate the H$\alpha$
  variability.  The $\sim$hour timescale optical dips that define the
  CPV class (blue arrows) tend to be contemporaneous with Balmer EW
  dips, consistent with transiting clumps of dusty plasma (see
  discussion in Section~\ref{sec:dustyplasma}).  The rightmost column
  is shown as a function of phase since the data span two orbital
  cycles.
  }
  \label{fig:dbsp}
\end{figure*}

\begin{figure*}[p]
  \centering
  \includegraphics[width=0.98\textwidth]{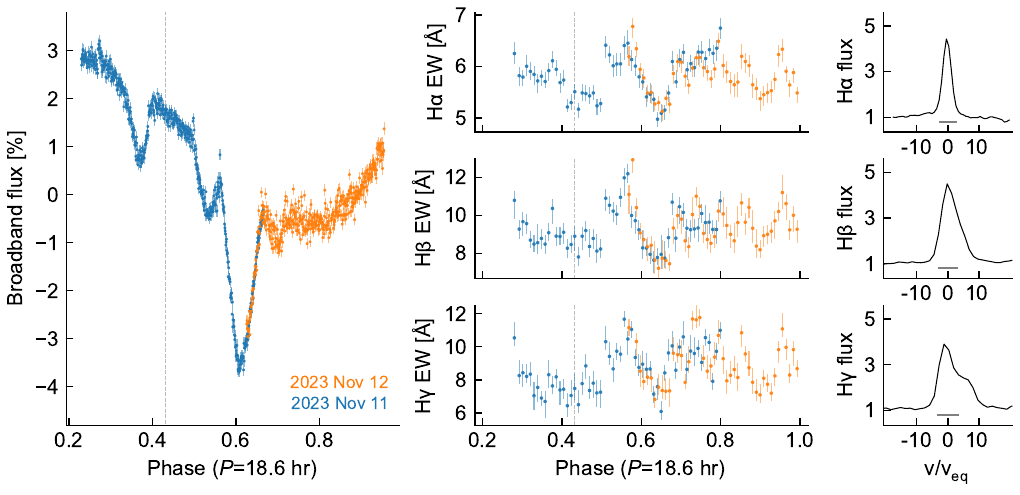}
  \vspace{-0.2cm}
  \caption{{\bf Hale/DBSP H$\alpha$, H$\beta$, and H$\gamma$
  equivalent widths and line profiles of LP~12-502, 11--12 November
  2023.}
  This figure details Figure~\ref{fig:dbsp}.  The
  \href{https://lgbouma.com/movies/LP_12-502_dbsp_Nov2023.mp4}{{\bf
  online movie}} shows the line profiles for each exposure, switching
  from Nov~11 to Nov~12 near $\phi$$\approx$0.8.  The target,
  LP~12-502, has $v_{\rm eq}$=24.2\,km/s; gray horizontal lines
  denote a single velocity resolution element
  ($\delta$$v$$\approx$$c$/R).  The Balmer line equivalent widths
  vary at the 5-50\% level, coincident in time with 1-2\% variability
  in the broadband Tierras filter.  There is strong evidence for line
  shape differences between H$\alpha$, H$\beta$, and H$\gamma$, and
  weak evidence for time variability in the velocity-resolved
  emission profile of H$\gamma$.  The jump at $\phi$$\approx$0.50 was
  likely flare-induced, based on similar timing coincidences in
  Figure~\ref{fig:dbsp}.}
  \label{fig:dbsp_detail}
\end{figure*}

The two LP~12-502 nights in November 2023 were conducted
simultaneously with the Tierras photometer at FLWO
\citep{Garcia-Mejia2020}, detailed in Section~\ref{sec:tierras} below.
The remaining observations, save for 7~December 2023, were conducted
simultaneously with TESS.  Figure~\ref{fig:dbsp} summarizes the
results from the most informative nights: 11~November to 12~November
2023, 4~November and 7~November 2024 for LP~12-502, and 28~April 2024
for DG~CVn.  Figure~\ref{fig:dbsp_detail} provides further detail on
the H$\alpha$, H$\beta$, and H$\gamma$ equivalent widths and line
profiles for the November 2023 nights.  Data from the remaining nights
of 5~November and 6~November 2024 for LP~12-502 are presented in
Appendix~\ref{app:ootspec}---these observations covered
orbital phases with no ``complex'' flux dips; only a single broadband
flare on 5~November yielded a corresponding spike in the Balmer lines
similar to those visible in Figure~\ref{fig:dbsp}.  The night of
7~December 2023 yielded viable DBSP data, but a TESS perigee contact
extended into the DBSP observation, leaving insufficient overlapping
photometry for further analysis.

% Nov05,Nov06 2024 - where just at an orbital phase that showed no flux dips

% Dec07 2023: poor TESS coverage due to orbit contact.  SAP_FLUX S73
% available starting BJDTDB 2460285.801.
% This only covers the latter third of the DBSP data, and the majority
% of variability in DBSP, likely including a large flare, happens
% before TESS data are available.

% DBSP LP12-502: extracted spectra showing 10-25k counts per pixel near
% Halpha continuum, depending on seeing and airmass

\subsection{Keck/HIRES}
\label{sec:hires}

We observed LP~12-502 ($V$=13.9) over 23~November~2023,
3~December~2024, and 10~December~2024 using the High Resolution
Echelle Spectrometer (HIRES; \citealt{vogt_hires_1994}) on the 10\,m
Keck I telescope.  We used the C2 decker
(0$\farcs$86$\times$14$\farcs$0) in the red configuration, yielding a
resolution $R$$\approx$45{,}000 ($\delta v$$\approx$6.7\,\kms).  The
seeing averaged 1$\farcs$2, 1$\farcs$6, and 2$\farcs$1 on the three
nights, respectively.
% S/N 40-70 per pixel; 3 pixels per resolution element along dispersion.
With 15-20 minute cadence, this configuration delivered an S/N of
70-120 per resolution element in the H$\alpha$ order's continuum, and
a total of 20.1\,hours of open-shutter time.  The
star's radius and period from \citet{Bouma2024} imply an equatorial
velocity (2$\pi$$R_\star$/$P$) of $v_{\rm eq}$=\lptwelveveq\,\kms, so
the instrument provided a few resolution elements across the line
core.  We reduced the echellogram using the standard techniques of the
California Planet Survey \citep{Howard2010}.  

We acquired simultaneous photometry for the December~2024 nights
from TESS.  We attempted to acquire
simultaneous Tierras data for the November~2023 night but
had poor weather in Arizona.
Appendix~\ref{app:ootspec} details the results.
In short, while the HIRES spectra and photometry of LP~12-502
presented abundant evidence for flare-induced variability, we observed
only a $\approx$one hour partial transit of a single ``complex'' dip,
which was coincident with excess absorption in Balmer line cores.

% NOTES:
% 23Nov2023: 05:41 - 13:17 UT j531.logsheet1, ~1.2", ~19k ct/exp, texp 1200, continuum near Ha shows 5k ct per px.
% 03Dec2024: 04:54 - 10:07 UT j546.logsheet2, ~1.6" windy, ~2.5k ct/exp, texp 900
% 10Dec2024: 04:28 - 11:48 UT j547.logsheet1, ~2.1" windy, ~2.5k ct/exp, texp 1200, continuum near Ha 2.8k ct per px

\subsection{Hale/TripleSpec}
\label{sec:tspec}

We observed LP~12-502 on 27~January~2024 (UT) using the Triple Spectrograph
(TripleSpec) mounted at the Cassegrain focus of the Palomar 200-inch
(5.1\,m) telescope.  TripleSpec is a medium-resolution,
cross-dispersed near-infrared spectrograph that provides simultaneous
spectral coverage from 1.0--2.4\,$\mu$m across five spectral orders
\citep{Herter2008}.   The instrument uses a $1\arcsec \times
30\arcsec$ slit and provides a resolving power of $R\sim2600$, with
approximately 2.7 pixels per resolution element.  The dispersed
spectrum is recorded using a HAWAII-II HgCdTe detector.
 
We collected 13 spectra of LP~12-502 between 02:00 and 06:15~UT, each
a 4$\times$120\,s ABBA sequence along the slit, allowing for the
removal of the near-infrared sky background, terrestrial OH emission,
and detector background through subtraction of adjacent nod positions.
We observed A0V telluric standards at similar airmass for telluric
correction and flux calibration.  We performed flat-field corrections,
wavelength calibration, spectral extraction, coaddition, telluric
correction, and merging of the individual spectral orders using the
IDL-based reduction package Spextool \citep{cushing2004}.

We modeled each spectrum independently following
\citet{PerezPaolino2025}, representing the flux as
\begin{equation}
  \begin{split}
    F_\lambda = \Big[ &(1-f_{\rm spot})\,F_\lambda(T_{\rm phot})
      + f_{\rm spot}\,F_\lambda(T_{\rm spot}) \\
      &+ f_{\rm excess}\,B_\lambda(T_{\rm excess}) \Big]
      \left(\frac{R_\star}{d}\right)^{\!2},
  \end{split}
  \label{eq:tspec}
\end{equation}
where $F_\lambda(T)$ are BT-Settl synthetic spectra \citep{Allard2014} for the unspotted
photosphere and the spots, $f_{\rm spot}$ is the spot filling factor,
and $B_\lambda$ is an optional blackbody with emitting area
$f_{\rm excess}$ relative to the visible stellar disk.  We sampled the
posteriors with \texttt{dynesty} \citep{Speagle2020} under uniform
priors, masked the telluric bands at 1.10--1.20, 1.325--1.45, and
1.80--2.05\,$\mu$m, and added a 3\% fractional model uncertainty in
quadrature with the measurement errors.  We repeated the fits with
the spot term removed, and separately with the excess term removed,
to assess whether the data required either.

\begin{figure}[htp!]
  \centering
  \includegraphics[width=0.49\textwidth]{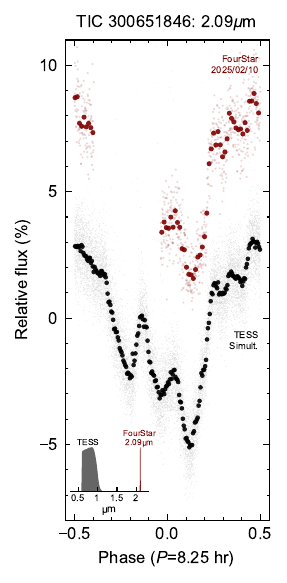}
  \vspace{-0.4cm}
	\caption{{\bf Magellan/FourStar 2.09\,$\mu$m photometry of
  TIC~300651846 vs.~simultaneous TESS (0.6--1\,$\mu$m).}  The target
  has a period of 8.25 hours and was observable for two-thirds of a
  rotation cycle with FourStar on 11~February~2025.  Simultaneous
  TESS data from Sector 88 are plotted underneath.  The two bands
  (filter profiles in lower left) show dip depths in a ratio of
  \fourstarobsratio$\pm$\fourstarobsratiounc\ (NB2.09/TESS).}
  \label{fig:fourstar}
\end{figure}

\begin{figure}[htp!]
  \centering
  \includegraphics[width=0.45\textwidth]{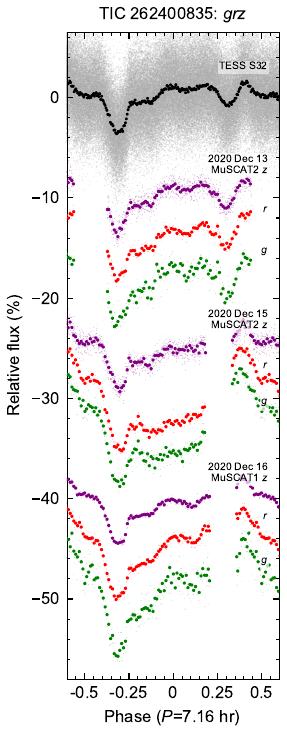}
  \vspace{-0.3cm}
  \caption{{\bf MuSCAT1 and MuSCAT2 $g'r'z_{\rm s}$ photometry of
  TIC~262400835 vs.~simultaneous TESS.}  The target has a period of
  7.16 hours.  The plotted TESS data, acquired at 20 second cadence,
  span 19 November 2020 to 17 December 2020.}
  \label{fig:muscat}
\end{figure}

\begin{figure}[!t]
  \centering
  \includegraphics[width=0.48\textwidth]{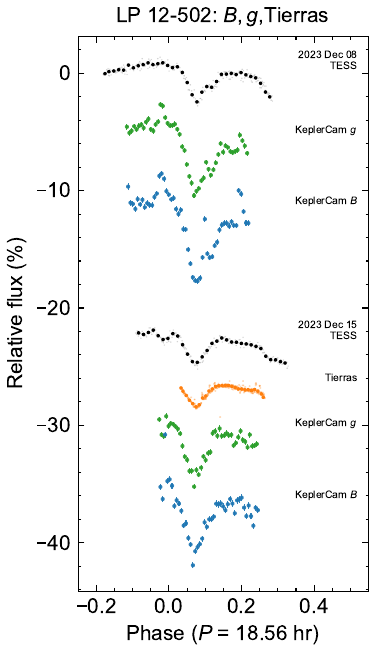}
  \vspace{-0.4cm}
  \caption{{\bf KeplerCam $gB$ and Tierras photometry of LP~12-502
  vs.~simultaneous TESS.}  The target has a period of 18.56 hours.
  The plotted TESS data are excerpts of 120 second cadence data from
  Sector~73.}
  \label{fig:keplercamtierras}
\end{figure}

\subsection{Magellan/FourStar}
\label{sec:fourstar}

We observed TIC~300651846 ($K$\,=\,10.8) on 11~February~2025 (UT) using
FourStar \citep{Persson2013}, a near-infrared wide-field imager on
the Baade 6.5\,m Magellan telescope.  We used the NB2.09 narrowband
filter ($\lambda$\,=\,2.082--2.118\,$\mu$m), in the FullWell read mode,
taking 20.38\,s correlated double sampled exposures.  The broader
$K$-band filter would have saturated the detector given the target's
brightness.  Young M dwarfs present no prominent emission or
absorption lines within the NB2.09 bandpass: the H$_2$ 1--0\,S(1)
transition at 21{,}218\,\AA\ falls off-band.  We therefore interpret
variability detected in this filter as continuum variability.

Humidity kept us closed over the first quarter of the night.  After
opening, we observed for 5.3\,hr beginning at BJD$_{\rm
TDB}$\,=\,2460717.650, with seeing averaging 0$\farcs$9.  We reduced
the images to light curves using {\tt AstroImageJ}
\citep[][]{Collins2017} and converted timestamps 
to BJD$_{\rm TDB}$ using the \citet{Eastman2010} barycentric
correction code.  Figure~\ref{fig:fourstar} shows the resulting light
curve phased to $t_0$\,=\,BJD$_{\rm TDB}$\,2460693.244 and
$P$\,=\,0.34392\,d (8.25\,hr), with the simultaneous TESS observation
from Sector~88.

\subsection{MuSCAT1 and MuSCAT2}
\label{sec:muscat}

We observed TIC~262400835 on 13 and 15~December~2020 UT using the
MuSCAT2 multi-band imager \citep{Narita2019} mounted on the 1.5\,m
Telescopio Carlos S\'{a}nchez (TCS) at Teide Observatory, Spain.
MuSCAT2 is equipped with four CCDs capable of obtaining simultaneous
imaging in the $g'$, $r'$, $i'$, and $z_{\rm s}$ bands with negligible
readout overhead. Each CCD consists of $1024 \times 1024$ pixels,
providing a field of view of $7.4 \times 7.4$\,\arcmin$^{2}$.

We conducted the MuSCAT2 observations with the telescope in nominal
focus. On both nights, we set the exposure times to 60\,s for the
$g'$, $r'$, and $i'$ bands, and 15\,s for the $z_s$ band.  We
processed the raw data using the MuSCAT2 pipeline
\citep{Parviainen2019}, which performs standard dark and flat-field
calibrations followed by aperture photometry.   We also used this
pipeline to model instrumental and environmental systematics,
specifically accounting for variations induced by airmass changes and
pixel-positional shifts due to telescope guiding.  We adopted
identical reference stars across both MuSCAT2 nights, and adopted an
aperture with $r$=16\,px ($7\farcs0$) to ensure negligible
contamination from TIC~262400836.  This aperture choice localized
the 7.16\,hr ``dip'' signal to TIC~262400835.

We also observed TIC~262400835 on the night of 16~December~2020 UT using
MuSCAT1 \citep{Narita2015}, a multi-band imager on 
the 1.88\,m telescope of the National Astronomical Observatory of Japan
(NAOJ) located in Okayama, Japan.
MuSCAT1 is equipped with three CCDs, enabling simultaneous imaging in
the $g'$, $r'$, and $z_s$ bands with negligible readout overhead. Each
CCD provides a field of view of $6.1 \times 6.1$\,arcmin$^2$ over
$1024 \times 1024$ pixels.

We used exposure times of 60 seconds in the $g'$, $r'$, and $z_{\rm s}$
bands.  We dark-subtracted and flat-fielded the raw images, and
performed aperture photometry following \citet{Fukui2011}, adopting an
aperture radii of 8, 10, and 14\,px ($2\farcs9$, $3\farcs6$, and
$5\farcs1$) to exclude the flux of the companion $\sim$10\arcsec\
to the west. These observations were conducted simultaneously with
TESS Sector 32, and the resulting light curves are shown in
Figure~\ref{fig:muscat} alongside the MuSCAT2 data.

\subsection{KeplerCam}
\label{sec:keplercam}

We observed LP~12-502 on 8 and 15~December 2023 UT using the KeplerCam
CCD on the 1.2\,m telescope at the Fred Lawrence Whipple Observatory
(FLWO) at Mount Hopkins, Arizona.   We took observations in
alternating $g'$ and $B$ filters with 60 sec exposure in $g'$ and 300
sec in $B$. KeplerCam is a 4096$\times$4096 Fairchild CCD which has a
field of view of $23\farcm1\times23\farcm1$ and an image scale of
0.672$"/$pix when binned by 2.   We reduced the data using standard IDL
routines, and performed aperture photometry using AstroImageJ
\citep{Collins2017}.
The resulting measurements are in Figure~\ref{fig:keplercamtierras}.
We additionally observed on the night of 16~December 2023, when high
winds and very poor seeing degraded the data quality.  The KeplerCam
$B$ and $g'$ light curves from that night weakly suggest the presence
of a flux dip, and the simultaneous TESS light curve shows a
$\approx$1\% deep dip lasting an hour.  Due to the poor quality of
these data relative to the nights of 8~December and 15~December, we
exclude them from further analysis.

\begin{deluxetable*}{llllcc}
\tablewidth{0pt}
\tabletypesize{\small}
\tablecaption{Multicolor Photometric Observations\label{tab:obs_phot}}
\tablehead{
  \colhead{Star} &
  \colhead{Instrument} &
  \colhead{Night} &
  \colhead{Bandpasses} &
  \colhead{Dips} &
  \colhead{Chromatic?} \\[-5pt]
  \colhead{} &
  \colhead{} &
  \colhead{(UT)} &
  \colhead{} &
  \colhead{Seen?} &
  \colhead{}
}
\startdata
TIC\,262400835 & MuSCAT2             & 2020 Dec 13 & $g'$, $r'$, $i'$, $z_{\rm s}$, TESS      & $\checkmark$ & ?            \\
TIC\,262400835 & MuSCAT2             & 2020 Dec 15 & $g'$, $r'$, $i'$, $z_{\rm s}$, TESS      & $\checkmark$ & $\checkmark$ \\
TIC\,262400835 & MuSCAT1             & 2020 Dec 16 & $g'$, $r'$, $z_{\rm s}$, TESS           & $\checkmark$ & $\checkmark$ \\
LP\,12-502     & KeplerCam           & 2023 Dec 08 & $B$, $g'$                      & $\checkmark$ & $\checkmark$ \\
LP\,12-502     & KeplerCam + Tierras & 2023 Dec 15 & $B$, $g'$, Tierras             & $\checkmark$ & $\checkmark$ \\
LP\,12-502     & KeplerCam           & 2023 Dec 16 & $B$, $g'$                      & $\checkmark$ & ? \\
TIC\,300651846 & FourStar + TESS     & 2025 Feb 11 & NB2.09, TESS                  & $\checkmark$ & ? \\
\enddata
\tablecomments{``Dips Seen?'' indicates whether sharp flux dips were
  observed; ``Chromatic?'' indicates whether the dip depths differed
  significantly between bandpasses.  The FourStar entry is the
  central near-infrared constraint of this work.}
\end{deluxetable*}

\subsection{Tierras}
\label{sec:tierras}

We observed LP\,12-502 and DG\,CVn with the Tierras Observatory, a
refurbished 1.3\,m ultra-precise fully-automated photometer located at
the F.~L.\ Whipple Observatory atop Mount Hopkins, Arizona
\citep{Garcia-Mejia2020}. The facility carries out all observations
through a single custom narrow bandpass filter ($\lambda_C =
863.5$~nm, FWHM $= 40$~nm) that minimizes precipitable water vapor
(PWV) errors known to limit photometric precision from the ground
\citep[e.g.][]{Stubbs2007,Blake2008,Berta2012}. Tierras is also
equipped with a deep-depletion $4\mathrm{K} \times 4\mathrm{K}$ e2v
CCD ($15~\mu$m pixels) with a plate scale of $0\farcs43$~pixel$^{-1}$
operated in frame-transfer (shutter-less) mode, yielding a field of
view (FOV) of $15' \times 29'$.

We observed LP\,12-502 during the UT nights of 11~November,
12~November, and 15~December 2023, as well as DG\,CVn on the UT night
of 28 April 2024.  The telescope was in focus throughout.  We exposed
for 45~s on LP\,12-502, collecting 556, 421, and 296 frames over 8.1,
6.1, and 4.2~hr, respectively, and for 30~s on DG\,CVn, collecting 731
frames over 7.9~hr. Our LP\,12-502 observations sampled airmasses
between 1.52 and 1.74, while our DG\,CVn observations sampled
airmasses between 1.00 and 1.80.
Skies were clear on all four nights, with relative humidities between
10\% and 52\%. The median image FWHM was $0\farcs9$, $1\farcs7$,
$1\farcs3$, and $1\farcs1$, listed chronologically.

To extract light curves from the Tierras images, we first
bias-corrected the images by subtracting bias levels derived from the
overscan region of each detector half.
%We did not flat-field the data given that, when these
%observations were gathered, light entering the Tierras optical path
%off-axis produced an additive glow that a flat field would not have
%corrected. This effect has since been suppressed with additional
%internal baffling.  
We then computed an astrometric solution for each
frame, and extracted photometry for stars with $G_\mathrm{RP}
< 17$.  We measured fluxes through a family of circular apertures
spanning radii of 5 to 20~pixels, estimating the local sky in each
case using annuli running from 35 to 55~pixels.  Our per-frame
error budget combined photon noise from the star and from the sky with
dark current, read noise, and scintillation \citep{Stefansson2017}.
We discarded frames in which the target flux encroached on the
detector's non-linear regime, which is why only 609 of the 731 DG\,CVn
frames were kept for the analysis.  We retained all 1273
LP\,12-502 frames.

Finally, we normalized our target fluxes by a weighted sum over the
ensemble of comparison stars, with the weights set by the iterative
scheme detailed in \citet{Tamburo2022}.  In this scheme, stars that
contribute noise are progressively downweighted until the weights stop
changing.  We then freeze the converged weights and reuse them for
every night of a given target. Holding the ensemble fixed in this way
guarantees that a brightness change measured for a variable target
between two nights cannot be an artifact of shifts in the comparison
set. For each target we kept the aperture that minimized the measured
scatter: 17~pixels ($7\farcs3$) for LP\,12-502, with an ensemble of
403 comparison stars, and 10~pixels ($4\farcs3$) for the brighter
DG\,CVn, with 86 comparison stars. Individual exposures for the two
targets reach a photometric precision of 1.1 to 1.7~ppt.

\subsection{Dip Depth Measurement}
\label{sec:dipfit}

We measured the depth, mid-time, and duration of each dimming event
in the ground-based photometry by treating
every observed dip independently.  We fitted the light curves using a
two-stage least squares routine.  We first manually provided a guess
for which points were ``in-dip'', which neighboring points provided
the ``out-of-dip'' baseline, and which points to exclude due to
flaring.  We constrained this initial guess to be identical across
bandpasses.  We then fit a smooth rotational baseline, $b(t)$, to only
the out-of-dip points, and subsequently fit the residual, $f(t) -
b(t)$, with a dip model.  We explored various models for the
rotational baseline (low-order polynomial; low-order Fourier series)
and the dip model (hyperbolic secant per \citealt{Yu2015}; trapezoidal
dip), always requiring the dip epoch and total duration to be
identical over bandpasses, but letting the depth and shape parameters
vary.  We assessed the goodness of fit for each model using the
reduced $\chi^2$, as initially evaluated by setting the flux
uncertainties to the point-to-point out-of-dip RMS (the 84$^{\rm
th}$-16$^{\rm th}$ percentile spread of the first differences $\Delta
F_j = F_j - F_{j+1}$), and subsequently renormalized to force the
best-fitting model to have $\chi^2_{\rm red} \equiv 1$.  We then
compared the model evidence using the Bayesian Information Criterion
(BIC).  We found across all the photometric datasets that
baseline models of at most second order minimized the BIC, and that the
trapezoidal dip model (parametrized by depth, duration, mid-time, and
fractional ingress/egress duration) offered better $\chired$
and $\Delta {\rm BIC}$ than the hyperbolic secant (parametrized by
depth, width, and mid-time).  Table~\ref{tab:dipdepths} lists the
resulting depth measurements, based on the trapezoidal model and the
BIC-preferred out-of-dip baseline models.
Appendix~\ref{app:fit_diagnostics} shows the resulting fits with each
photometric dataset.

% Single-column compilation of measured dip depths vs.~bandpass.
% Data rows are auto-generated by `python -m fit_dips.make_table`
% between the BEGIN/END sentinels; the \tablecomments are hand-kept.
\begin{deluxetable}{lllcl}
\tabletypesize{\scriptsize}
\tablewidth{0pt}
\tablecaption{Dip depths by bandpass\label{tab:dipdepths}}
\tablehead{
  \colhead{Star} &
  \colhead{Date} &
  \colhead{Band} &
  \colhead{$\delta$ (\%)} &
  \colhead{Ref.} \\[-5pt]
  \colhead{} &
  \colhead{(UT)} &
  \colhead{} &
  \colhead{} &
  \colhead{}
}
\startdata
%\cutinhead{This work}
% BEGIN fit-dips rows -- auto-generated by `python -m fit_dips.make_table`; do not edit by hand.
TIC\,262400835 & 2020 Dec 13    & $g'$        & $3.36\pm 0.19$ & 1 \\
               &                & $r'$        & $2.54\pm 0.10$ & 1 \\
               &                & $i'$        & $2.74\pm 0.07$ & 1 \\
               &                & $z_{\rm s}$ & $2.69\pm 0.09$ & 1 \\
               &                & TESS$^{\ddagger}$ & $0.96\pm 0.12$ & 1 \\
               & 2020 Dec 15    & $g'$        & $4.67\pm 0.15$ & 1 \\
               &                & $r'$        & $3.46\pm 0.10$ & 1 \\
               &                & $i'$        & $3.64\pm 0.08$ & 1 \\
               &                & $z_{\rm s}$ & $2.99\pm 0.09$ & 1 \\
               &                & TESS$^{\ddagger}$ & $1.95\pm 0.23$ & 1 \\
               & 2020 Dec 16    & $g'$        & $4.30\pm 0.20$ & 1 \\
               &                & $r'$        & $3.52\pm 0.08$ & 1 \\
               &                & $z_{\rm s}$ & $2.96\pm 0.06$ & 1 \\
               &                & TESS$^{\ddagger}$ & $2.17\pm 0.38$ & 1 \\
LP\,12-502     & 2023 Dec 08    & $B$         & $5.57\pm 0.28$ & 1 \\
               &                & $g'$        & $4.63\pm 0.18$ & 1 \\
               &                & TESS        & $2.44\pm 0.08$ & 1 \\
               & 2023 Dec 15    & $B$         & $4.48\pm 0.44$ & 1 \\
               &                & $g'$        & $3.48\pm 0.24$ & 1 \\
               &                & Tierras     & $2.75\pm 0.09$ & 1 \\
               &                & TESS        & $2.18\pm 0.24$ & 1 \\
TIC\,300651846 & 2025 Feb 11    & TESS        & $4.07\pm 0.05$ & 1 \\
               & (dip 1)        & NB2.09      & $3.27\pm 0.10$ & 1 \\
               & (+1.4 h)       & TESS        & $0.62\pm 0.08$ & 1 \\
               &                & NB2.09      & $1.06\pm 0.20$ & 1 \\
               & (+2.1 h)       & TESS        & $0.80\pm 0.05$ & 1 \\
               &                & NB2.09      & $0.81\pm 0.10$ & 1 \\
TIC\,435899024 & 2021 May 25--  & $B$         & $12.47\pm 0.40$ & 2$^\dagger$ \\
               & --2021 June 01 & $V$         & $11.96\pm 0.41$ & 2$^\dagger$ \\
               &                & $R$         & $10.79\pm 0.29$ & 2$^\dagger$ \\
               &                & $I$         & $9.11\pm 0.21$ & 2$^\dagger$ \\
PTFO\,8-8695   & 2014 Jan 19    & $I{+}z$     & $2.18\pm 0.08$ & 3$^\dagger$ \\
               &                & $H$         & $1.38\pm 0.05$ & 3$^\dagger$ \\
               & 2016 Feb 9     & $g'$        & $2.57\pm 0.34$ & 4$^\dagger$ \\
               &                & $r'$        & $1.69\pm 0.12$ & 4$^\dagger$ \\
               &                & $z_{\rm s}$ & $0.99\pm 0.10$ & 4$^\dagger$ \\
               & 2014 Feb 23    & $I_{\rm C}$ & $2.59\pm 0.10$ & 5$^\dagger$ \\
               &                & $K_{\rm s}$ & $1.07\pm 0.20$ & 5$^\dagger$ \\
               & 2014 Dec 27    & $I_{\rm C}$ & $1.25\pm 0.08$ & 5$^\dagger$ \\
               &                & $K_{\rm s}$ & $0.97\pm 0.11$ & 5$^\dagger$ \\
               & 2015 Jan 10    & $I_{\rm C}$ & $1.34\pm 0.13$ & 5$^\dagger$ \\
               &                & $K_{\rm s}$ & $1.24\pm 0.22$ & 5$^\dagger$ \\
               & 2015 Jan 23    & $I_{\rm C}$ & $1.56\pm 0.14$ & 5$^\dagger$ \\
               &                & $J$         & $1.00\pm 0.22$ & 5$^\dagger$ \\
               & 2015 Feb 10    & $I_{\rm C}$ & $1.64\pm 0.15$ & 5$^\dagger$ \\
               &                & $H$         & $0.87\pm 0.11$ & 5$^\dagger$ \\
               & 2015 Feb 14    & $I_{\rm C}$ & $2.16\pm 0.05$ & 5$^\dagger$ \\
               &                & $K_{\rm s}$ & $1.10\pm 0.10$ & 5$^\dagger$ \\
               & 2015 Feb 23    & $I_{\rm C}$ & $1.44\pm 0.17$ & 5$^\dagger$ \\
               &                & $K_{\rm s}$ & $1.12\pm 0.27$ & 5$^\dagger$ \\
               & 2016 Nov 25    & $I_{\rm C}$ & $1.01\pm 0.12$ & 5$^\dagger$ \\
               &                & $J$         & $1.09\pm 0.11$ & 5$^\dagger$ \\
               & 2016 Dec 9     & $I_{\rm C}$ & $1.38\pm 0.15$ & 5$^\dagger$ \\
               &                & $J$         & $0.90\pm 0.10$ & 5$^\dagger$ \\
% END fit-dips rows
\enddata
\tablecomments{$^\dagger$Denotes literature photometry refitted in
  this work.  $^{\ddagger}$Listed for completeness but excluded from
  the extinction law fits due to low S/N and an uncertain crowding
  correction (Section~\ref{subsec:extinction_methods}).
  {\it References}: (1)  This work.
  (2) \citet{Koen2023} SAAO\,1m/CCD ``J1252'';
  (3) \citet{Yu2015} Baade\,6.5m/FourStar $H$ and TRAPPIST $I{+}z$;
  (4) \citet{Onitsuka2017} NAOJ\,1.88\,m/MuSCAT simultaneous;
  (5) \citet{Tanimoto2020} HHO\,1.5\,m/HONIR.}
\end{deluxetable}

\section{Interpretation}
\label{sec:results}

\subsection{Bound circumstellar gas in emission}
\label{subsec:mage_interpret}

% Paragraph #1: MagE TIC300651846 shows that TIC1411 was not a fluke.
The 18~January~2025 Magellan/MagE observations of TIC~300651846
(Figure~\ref{fig:mage}) show sinusoidal time-varying H$\alpha$
emission with semi-amplitude $K$$\approx$4\,$v_{\rm eq}$.  To our
best knowledge, the only available explanation is emission from a cool
($\lesssim$10$^4$\,K) plasma clump trapped in corotation.  This is the
second CPV showing this phenomenology, after TIC~141146667
\citep{Bouma2025}.  TIC~141146667 showed two clumps at opposite
stellar longitudes, which is an azimuthal distribution predicted by
models of rigidly rotating magnetospheres with tilted dipolar magnetic
fields \citep{Townsend2008}.  Such models tend to produce ``pinched''
plasma tori\footnote{See
\url{http://user.astro.wisc.edu/\~townsend/static.php?ref=rrm-movies}
last accessed 26 Aug 2026.}.  For TIC~300651846, the clearest signal
shows a single clump appearing at $\phi$=0.1 near
$\approx$4\,$R_\star$ (beyond $R_{\rm c}$=2.49$\pm$0.15\,$R_\star$), and
rotating with the star for the next six hours.  The clump emits with an EW
$\approx$60\% that of the star, and is internally unresolved at MagE's
resolving power ($\approx$73\,\kms).  A secondary feature in the
18~January~2025 night appears from $\phi$=0.6-0.7.  While this feature's
interpretation is complicated by its brief lifetime, the star presents a
triple-lined H$\alpha$ profile in three separate exposures.  This is consistent
with the idea of opposing plasma clumps becoming visible at phase quadrature,
when they are maximally separated in velocity.

We estimate the number density of the longer-lived plasma clump based
on the emitting line's luminosity as follows.  For the star, we
observe a time-averaged H$\alpha$ equivalent width of $6.0 \pm
0.3$\,\AA.  Given the SED-derived $L_\star$=0.033$\pm$0.006\,L$_\odot$
\citep{Bouma2024}, this implies that the stellar H$\alpha$ line
radiates $\approx$3.0$\times$10$^{28}$\,erg\,s$^{-1}$.  From the same
SED fit, the star has $R_\star$=0.61$\pm$0.04\,R$_\odot$.  Following
Appendix~C of \citet{Bouma2025} and approximating all the emission as
coming from case B radiative recombination, a homogeneous sphere of
radius $r$$\approx$0.1\,$R_\star$ would yield $n_{\rm
H}$$\approx$8$\times$10$^{11}$\,cm$^{-3}$, with a total clump mass
$M_{\rm gas}$$\approx$5$\times$10$^{17}$\,g---all on the night of
18~January~2025.

The other nights in January~2025 showed much weaker circumstellar
emission.  Close inspection of the 19~January and 20~January MagE data
(Figure~\ref{fig:mage}) shows an emissive component appearing at
similar orbital phases as on 18~January.  The MIKE data from 2~January
show a similar signal (Figure~\ref{fig:mike}).  However, this emission
has an equivalent
width $\approx$10\% that of the star's average
H$\alpha$ profile, rather than $\approx$60\%; on 21~January, no
emission is detectable.  Under radiative recombination, $n_{\rm H}
\propto L_{\rm H\alpha}^{1/2}$, so this six-fold decrease in
luminosity corresponds to a number density decrease of 2.5$\times$.
Despite this change in the circumstellar plasma number density, the
simultaneous light curve morphology in Figure~\ref{fig:mage} appears
unchanged.

The implication is that, at least over these five nights of
observations, the light curve morphology was not directly tied to the
circumstellar plasma number density.  This is consistent with previous
discussion \citep[e.g.~Section 5.6 of][]{Bouma2024} arguing that there
is no microphysical mechanism by which gas can provide the opacity
needed to explain both the observed transit depths and their
chromaticities.
The spectra show that gas clumps are present---but dust remains the
sole viable opacity source.  The gas and dust broadly trace each
other, in that Balmer dimming accompanies the broadband dips
(Section~\ref{subsec:ha_flux_dip_interpret}).  However, the large
swings in gas emissivity without any change in dip morphology
indicate that the dust column is not simply proportional to the gas
density.

\subsection{Balmer dimming during transits}
\label{subsec:ha_flux_dip_interpret}

With lower resolution spectra, it is no longer possible to separate
circumstellar emission from the stellar chromosphere.
Nonetheless, when a plasma clump transits, one would expect the
plasma to scatter or absorb stellar H$\alpha$ photons, yielding an
H$\alpha$ transit that mirrors the broadband flux.  We refer to this
combination of absorption and scattering ({\it i.e.}, extinction) as dimming.
Figures~\ref{fig:dbsp} and~\ref{fig:dbsp_detail} show low-resolution
data that appear consistent with the hypothesis of greater H$\alpha$
dimming during broadband transits: the local minima in flux are
correlated with local minima in the H$\alpha$ equivalent width
time-series.  Moreover, Figures~\ref{fig:mage} and~\ref{fig:mike} both
show similar absorption in the H$\alpha$ line cores of TIC~300651846
from $\phi$=0.9--1.0, consistent with absorption or scattering from
the same plasma clump that is then seen in emission at phase
quadrature.

This type of low
resolution data presents two difficulties.  The first is flares:
minute-timescale optical flares produce
large-amplitude, hour-long
Balmer flares.  The orange arrows in Figure~\ref{fig:dbsp} and
in Appendix~\ref{app:dbsp_ootspec} highlight such flare-induced
variability, which complicates interpretation of H$\alpha$ EW
time-series data.  The second challenge is connected to the {\it
consistency} with which local minima in flux yield local minima in
H$\alpha$ EW.  The broadband flux dips in the three leftmost columns
of Figure~\ref{fig:dbsp} are all coincident in time with local minima
in their corresponding H$\alpha$ EW time-series.  However, the nights
of 11 and 12 November 2023 showed four local minima in flux.  The
largest of these local flux minima, which was observed on both nights
at $\phi$$\approx$0.62, corresponds to an H$\alpha$ minimum
that {\it lagged} the flux minimum on both nights by
$\approx$50 minutes.  The green ``H$\alpha$ core'' lines in
Figures~\ref{fig:mage} and~\ref{fig:mike} show an analogous offset
with the opposite sign: for TIC~300651846, the deepest H$\alpha$
minimum {\it led} the deepest broadband flux minimum by
$\approx$25 minutes on all five nights in January~2025.  While the origin
of these time offsets is not clear, their repeatability suggests
that they are a stable feature associated with the material that
causes CPV dips.

Summarizing, the low-resolution observations show
$\delta$$\approx$1-3\% broadband flux dips corresponding to
$\approx$5-20\% variations in H$\alpha$ EW.  If the occulting cloud
were gas-free, and made only of dust, this would not happen: dust does
not selectively absorb light at 6563\,\AA.  These observations are
therefore consistent with the idea that the transiting material must
include hydrogen.

\setlength{\tabcolsep}{3pt}
\begin{deluxetable*}{lcccccccc}
\tablewidth{0pt}
\tabletypesize{\footnotesize}
  \tablecaption{Extinction law fitting results ($\delta \propto
  \lambda^{-\beta}$), including goodness of
  fit and the mean $\pm$ std.~dev.~for $\beta$.  The
  number of model free parameters $k$ and degrees of freedom $\nu=N-k$
  are given as no-$\sint$/with-$\sint$ pairs. Both $\chired$ columns
  are evaluated against measurement error alone, $s_i$, not the value
  $\approx$$1$ obtained once the fitted $\sint$ is added in quadrature. 
  $\dBIC$ is relative to the per-epoch model within each variant.
\label{tab:fitresults}}
\tablehead{
\colhead{} & \colhead{} & \colhead{} &
\multicolumn{2}{c}{$\chired$} &
\multicolumn{2}{c}{$\dBIC$} &
\multicolumn{2}{c}{$\beta$} \\
\cline{4-5} \cline{6-7} \cline{8-9}
\colhead{Model} & \colhead{$k$} & \colhead{$\nu$} &
\colhead{no $\sint$} & \colhead{with $\sint$} &
\colhead{no $\sint$} & \colhead{with $\sint$} &
\colhead{no $\sint$} & \colhead{with $\sint$}
}
\startdata
%
% 2026/07/04: copied from ../../results/beta_fit/combined/tab_fitresults_rows.tex
%
Single & 21/22 & 30/29 & 8.39 & 13.85 & 127.8 & 23.0 & $0.47\pm0.02$ & $0.46\pm0.08$ \\[3pt]
Broken \tablenotemark{a} & 23/24 & 28/27 & 7.25 & 12.31 & 76.6 & 16.3 & --- & --- \\
\quad $\beta_1$ (blueward) & --- & --- & --- & --- & --- & --- & $0.68\pm0.07$ & $0.79\pm0.16$ \\
\quad $\beta_2$ (redward) & --- & --- & --- & --- & --- & --- & $-0.10\pm0.31$ & $-0.09\pm0.39$ \\[3pt]
Per-star & 25/26 & 26/25 & 5.43 & 10.58 & 32.9 & 7.7 & --- & --- \\
\quad TIC 262400835 & --- & --- & --- & --- & --- & --- & $0.51\pm0.05$ & $0.51\pm0.18$ \\
\quad LP 12-502 & --- & --- & --- & --- & --- & --- & $0.98\pm0.07$ & $0.97\pm0.21$ \\
\quad TIC 300651846 & --- & --- & --- & --- & --- & --- & $0.19\pm0.04$ & $-0.01\pm0.14$ \\
\quad TIC 435899024 & --- & --- & --- & --- & --- & --- & $0.55\pm0.06$ & $0.53\pm0.32$ \\
\quad PTFO 8-8695 & --- & --- & --- & --- & --- & --- & $0.59\pm0.05$ & $0.57\pm0.10$ \\[3pt]
Per-epoch & 40/41 & 11/10 & 4.48 & 5.79 & 0.0\tablenotemark{b} & 0.0\tablenotemark{b} & --- & --- \\
\quad TIC 262400835, 2020 Dec 13 & --- & --- & --- & --- & --- & --- & $0.21\pm0.10$ & $0.28\pm0.21$ \\
\qquad 2020 Dec 15 & --- & --- & --- & --- & --- & --- & $0.60\pm0.07$ & $0.61\pm0.19$ \\
\qquad 2020 Dec 16 & --- & --- & --- & --- & --- & --- & $0.60\pm0.09$ & $0.61\pm0.23$ \\
\quad LP 12-502, 2023 Dec 08 & --- & --- & --- & --- & --- & --- & $1.32\pm0.09$ & $1.34\pm0.21$ \\
\qquad 2023 Dec 15 & --- & --- & --- & --- & --- & --- & $0.54\pm0.11$ & $0.65\pm0.19$ \\
\quad TIC 300651846, epoch 1 & --- & --- & --- & --- & --- & --- & $0.22\pm0.04$ & $0.22\pm0.13$ \\
\qquad epoch 2 & --- & --- & --- & --- & --- & --- & $-0.54\pm0.25$ & $-0.54\pm0.26$ \\
\qquad epoch 3 & --- & --- & --- & --- & --- & --- & $-0.01\pm0.18$ & $-0.01\pm0.19$ \\
\quad TIC 435899024, 2021 May 25--Jun 01 & --- & --- & --- & --- & --- & --- & $0.55\pm0.06$ & $0.53\pm0.20$ \\
\quad PTFO 8-8695, 2014 Jan 19 & --- & --- & --- & --- & --- & --- & $0.63\pm0.12$ & $0.63\pm0.19$ \\
\qquad 2014 Feb 23 & --- & --- & --- & --- & --- & --- & $0.87\pm0.19$ & $0.87\pm0.22$ \\
\qquad 2014 Dec 27 & --- & --- & --- & --- & --- & --- & $0.25\pm0.14$ & $0.25\pm0.18$ \\
\qquad 2015 Jan 10 & --- & --- & --- & --- & --- & --- & $0.08\pm0.20$ & $0.08\pm0.23$ \\
\qquad 2015 Jan 23 & --- & --- & --- & --- & --- & --- & $0.96\pm0.50$ & $0.94\pm0.58$ \\
\qquad 2015 Feb 10 & --- & --- & --- & --- & --- & --- & $0.85\pm0.21$ & $0.85\pm0.26$ \\
\qquad 2015 Feb 14 & --- & --- & --- & --- & --- & --- & $0.67\pm0.10$ & $0.67\pm0.15$ \\
\qquad 2015 Feb 23 & --- & --- & --- & --- & --- & --- & $0.26\pm0.26$ & $0.25\pm0.29$ \\
\qquad 2016 Feb 9 & --- & --- & --- & --- & --- & --- & $1.50\pm0.25$ & $1.50\pm0.32$ \\
\qquad 2016 Nov 25 & --- & --- & --- & --- & --- & --- & $-0.16\pm0.34$ & $-0.16\pm0.43$ \\
\qquad 2016 Dec 9 & --- & --- & --- & --- & --- & --- & $0.93\pm0.34$ & $0.93\pm0.43$ \\
\enddata
%
% end copy
%
\tablenotetext{a}{Break wavelength
$\lambda_b=\brokenlambdabreaknosig\pm\brokenlambdabreaknosigunc\um$ (no
$\sint$), $\brokenlambdabreak\pm\brokenlambdabreakunc\um$ (with $\sint$).}
\tablenotetext{b}{All four models suffer pathologies connected
to the need to fit the reference depth terms ``$\delta_{0,i}$''.
However, this problem is most severe for the per-epoch model, since 13
of 20 epochs have two bands and are therefore fit exactly (zero
residual $\chi^2$ regardless of the true $\beta$).  This model's
minimization of $\dBIC$ is an artifact of the incidental-parameters
problem, not evidence of real epoch-to-epoch variation.  See
Section~\ref{subsec:extinction_results}.}
\end{deluxetable*}

\subsection{A near-infrared break in the extinction law}
\label{subsec:beta}

CPV dips tend to appear deeper in the blue than in the red
\citep{Gunther2022,Koen2023}, which \citet{Koen2023} attributed to
extinction by sub-micron dust.   
Since methodological choices in  detrending and dip fitting 
can affect the inferred chromaticity, we opted to measure the dip depths from
both our new light curves and archival photometry using a single procedure.

\subsubsection{Extinction law methodology}
\label{subsec:extinction_methods}

\begin{figure*}[!t]
  \centering
  \includegraphics[width=0.99\textwidth]{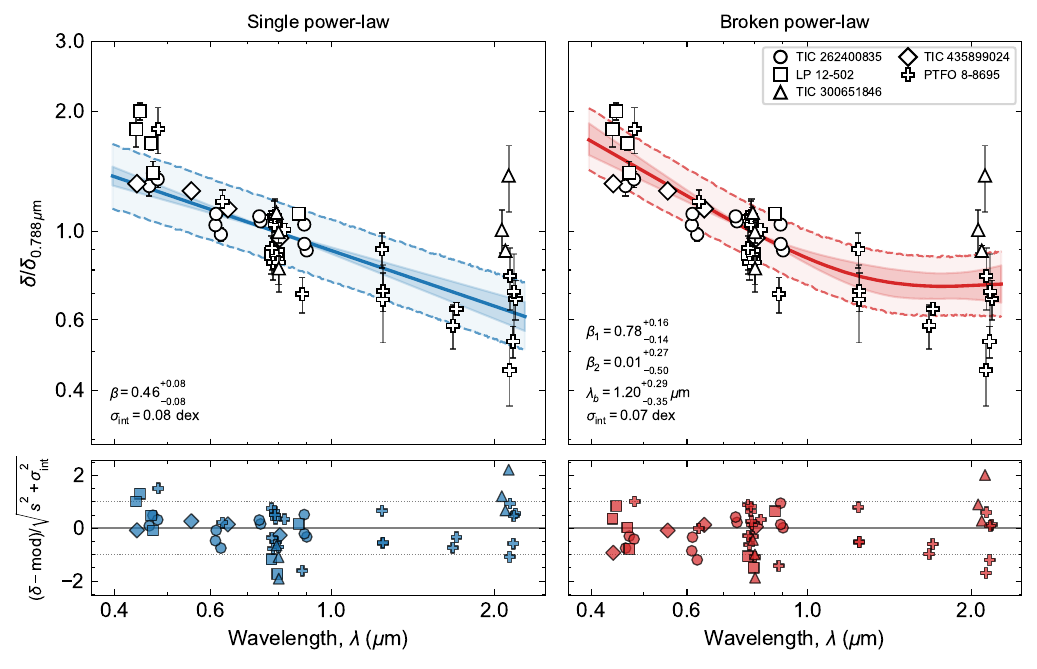}
  \vspace{-0.3cm}
  \caption{{\bf CPV dips are chromatic, with a near-infrared break.}
  Markers show the measured dip depth in each bandpass from
  Table~\ref{tab:dipdepths}, with small horizontal jitter added for
  visual clarity.  The normalization factor
  $\delta_{0.788\,\mu{\rm m}}$, the depth at TESS-band
  ($\lambda$=0.788\,$\mu$m), is a model-derived quantity per
  combination of star and observing epoch (see
  Section~\ref{subsec:beta}).  The solid line and inner band are the
  posterior-median and the 68\% confidence interval of the mean
  $\delta(\lambda)$ relation; the outer band is the 68\% predictive
  interval after additionally accounting for intrinsic variability
  ($\sigma_{\rm int}$).  Including $\sigma_{\rm int}$, there is a
  positive preference ($\Delta$BIC=\dbicsinglebroken) in favor of the
  broken power-law.  Excluding $\sigma_{\rm int}$, there is a strong
  preference for the broken power-law
  ($\Delta$BIC=\dbicsinglebrokennosig).  {\it References}:
  TIC\,262400835, LP\,12-502, and TIC\,300651846: this work.
  TIC\,435899024: \citet{Koen2023}.  PTFO\,8-8695: \citet{Yu2015},
  \citet{Onitsuka2017}, \citet{Tanimoto2020}.  }
  \label{fig:beta}
\end{figure*}

We collected literature photometry from observations of CPVs showing
at least one dip in at least two simultaneous bandpasses.
We required the dips to be
visually distinguishable (based on d$f$/d$t$) from spot-induced
modulation, which is also chromatic.  We excluded non-simultaneous
literature datasets even when they carried novel information
(e.g.~the Spitzer 4.5\,$\mu$m light curve from \citealt{Ciardi2015})
due to the time-variable nature of CPV dip depths.  
Further, we excluded {\it (i)} the TESS-band measurements for
TIC~262400835, due to their low S/N and uncertain crowding correction;
and {\it (ii)} the simultaneous SPECULOOS-South and TESS photometry of
TIC~20178925 from \citet{Gunther2022}, because its ground-based and TESS
dips show different durations, which is inconsistent with the fitting
assumptions we describe below.  

Table~\ref{tab:dipdepths} lists the results:
51 dip depth measurements spanning
0.44--2.16\,$\mu$m over five stars and twenty epochs.
Half of these measurements are new from this work; the remainder are
from TIC~435899024 \citep{Koen2023} and PTFO~8-8695
\citep{Onitsuka2017,Tanimoto2020}.

We fitted an extinction law to these dip depths by considering
four variants of a model in which
\begin{equation}
  \delta \propto \lambda^{-\beta},
\end{equation}
where $\delta$ is the dip depth and $\beta$ encodes how strongly the
occultation deepens toward the blue.  Positive $\beta$ corresponds to
deeper dips at shorter wavelengths, as expected for extinction by
sub-micron dust; $\beta=0$ is achromatic (gray), as expected for an
optically thick occulter.

Working in logarithmic space, we defined $y_i = \log_{10} \delta_i$,
yielding an uncertainty $\sigma_y \approx |{\rm d}y/{\rm d}{\delta}|
\sigma_\delta $ = $\sigma_{\delta} / (\delta \ln 10)$, for dip index
$i$ at an epoch $e(i)$.  Here ``$e(i)$'' refers to the set of
all multi-wavelength measurements of one dip at a given epoch, and 
``$s(i)$'' refers to the set of all observed dips for any given
star.  We further found it convenient to define $\Lambda_i \equiv
\log_{10}(\lambda_{{\rm eff},i}/ \lambda_0)$, where $\lambda_0$ is an
arbitrary reference wavelength, with a corresponding depth at that
wavelength $\delta_0$.  We chose
$\lambda_0$=0.788\,$\mu$m, the effective wavelength of the TESS
bandpass, and calculated the effective wavelengths of the various
bandpasses using the SVO Filter Profile Service \citep{Rodrigo2020}.
We then fitted the following four models to the dip depths in
Table~\ref{tab:dipdepths}:
\begin{align}
  \mu_i &=  c_{e(i)} - \beta \Lambda_i  \quad {\rm (single\ power\ law)}, \label{eq:single} \\
  \mu_i &=  c_{e(i)} - \beta_{s(i)} \Lambda_i  \quad {\rm (per-star\ power\ law)}, \\
  \mu_i &=  c_{e(i)} - \beta_{e(i)} \Lambda_i  \quad {\rm (per-epoch\ power\ law)}, \\
  \mu_i &=  c_{e(i)} - \beta_1 w_i + (\beta_1 - \beta_2) h_i \quad {\rm (broken\ power\ law)}, \label{eq:broken}
\end{align}
where $c_{e(i)} \equiv \log_{10} \delta_{0,e(i)}$ is a fitted reference
depth defined at the reference wavelength $\lambda_0$.  For observations that
have TESS measurements, $c_{e(i)}$ will be near the observed TESS
depth; for those without, $c_{e(i)}$ will be an extrapolation based on
whichever bands were observed.  In the broken power-law
model, $w_i \equiv \log_{10}(\lambda_{{\rm eff},i}/\lambda_b)$ is the
log-wavelength offset from a free break wavelength $\lambda_b$, and
$h_i \equiv \kappa \ln(1 + e^{w_i/\kappa})$ is a smoothing function
that ensures a continuous gradient in the transition from $\beta_1$
blueward of the break to $\beta_2$ redward of it.  We
take $\lambda_b$ as the reference wavelength for
the per-epoch amplitudes $c_{e(i)}$ rather than $\lambda_0$, and we
set the smoothing scale to $\kappa$=0.01\,dex, since our data do not
constrain it.  Finally, our priors on the fits were $\beta \sim
\mathcal{N}(0,3)$, $c_{e(i)} \sim \mathcal{N}(\langle y_i \rangle,
1)$, and, for the broken power-law model, $\lambda_b/\mu{\rm m} \sim
\mathcal{U}(0.7,1.6)$.  We assumed a Gaussian likelihood, $y_i \sim
\mathcal{N}(\mu_i,\sigma_i)$, where we let $\sigma_i = (s_i^2 +
\sigma_{\rm int}^2)^{1/2}$ optionally add an intrinsic scatter term to
the per-dip uncertainties.  We sampled using PyMC \citep{pymc2023},
and required $\hat{R} < 1.01$ with no divergences as an indicator of
convergence.

\subsubsection{Extinction law results}
\label{subsec:extinction_results}

Figure~\ref{fig:beta} and Table~\ref{tab:fitresults} summarize the
results.  While Figure~\ref{fig:beta} shows the single power-law and
broken power-law results in the $\delta$/$\delta_{0.788\,\mu{\rm
m}}$--$\lambda_{\rm eff}$ plane, we emphasize that the fits are in the
space of $y_i$ vs.~$\Lambda_i$.  The normalization
$\delta_{0.788\,\mu{\rm m}}$ is the model-derived reference depth at
the TESS effective wavelength for each dip epoch in
Equations~\ref{eq:single}--\ref{eq:broken}.  To provide a consistent
normalization for visual comparison, we adopted the
$\delta_{0.788\,\mu{\rm m},i}$ values from the ``per-star'' power law
fits, a choice that favors neither of the two displayed models.

All four models agree that the dip depths are chromatic.  The model
comparison diagnostics also indicate that a single power-law
extinction curve can be improved upon, either by introducing a
population-wide flattening of the extinction toward the near-infrared,
or by invoking epoch-to-epoch or star-to-star variability in the
power-law index.  Key issues in the model comparison include whether
the dip depth uncertainties $s_i$ are reliable; how to weigh each
model's parameter count; and the degree to which the data can select
between the models.

If the dip depth uncertainties were unreliable, then we would want to
incorporate an intrinsic scatter term $\sigma_{\rm int}$, effectively
down-weighting the most precise measurements and weakening our ability
to make model selection claims.
We calibrated the reported depth uncertainties as described in
Section~\ref{sec:dipfit}, using the empirical out-of-dip scatter in
each time-series dataset.  The data and fitted model uncertainties (see
Appendix~\ref{app:fit_diagnostics}) suggest that this procedure 
indeed yields uncertainties that match the scatter in the data.
For example, our refit to the archival photometry from
\citet{Onitsuka2017} yielded statistical uncertainties 10-17$\times$
larger than the published values, mirroring the observed scatter.  Our
dip depth uncertainties therefore seem plausible.

Nonetheless, there is an issue with treating our dip depth
uncertainties at face value: both the universal power-law and the
broken power-law models provide poor fits to the data
($\chi^2_\nu$=\singlechirednosig\ and \brokenchirednosig,
respectively).  While inspection of $\chi^2_\nu$ and $\Delta {\rm
BIC}$ in Table~\ref{tab:fitresults} would suggest that letting $\beta$
vary on a per-epoch basis is preferable, that alternative also comes
with drawbacks.  In particular, the ``$\beta$ per-epoch'' model is
pathological for dips observed with only two simultaneous bandpasses,
comprising over half the dataset: this model precludes $\chi^2$ or BIC
from accruing any residual information because it adds free parameters
at the same rate as new data points. 
%
%While we explored leave-one-out
%style analyses as a possible replacement strategy for counting {\it
%effective} free parameters, 
%
Ultimately, we found that with this dataset we were unable to
produce a reliable ranking of this ``per-epoch'' model compared against
the other models.  To some
degree, all four models share this ``incidental parameters problem''
\citep{Lancaster2000}, in that every new observation epoch requires
fitting and marginalizing over the depth at the reference wavelength;
the problem is just most severe for the per-epoch model.

In brief, two divergent options can improve on the single power law:
one can introduce a population-wide break that flattens the
extinction toward the near-infrared, or one can
invoke epoch-to-epoch or star-to-star variability.  
Assuming that the extinction law is universal,
the preferred model is a broken power-law
with $\beta_1 = \brokenbetaone \pm \brokenbetaoneunc$ blueward of the
break at $\lambda_b = \brokenlambdabreak \pm
\brokenlambdabreakunc$\,$\mu$m, and $\beta_2 = \brokenbetatwo \pm
\brokenbetatwounc$ redward of it.  Intuitively,
this model seems to provide the most plausible explanation for the narrowband
FourStar 2.09\,$\mu$m observations shown in Figure~\ref{fig:fourstar},
which yielded dip depths
\fourstarobsratio$\pm$\fourstarobsratiounc$\times$ the TESS depth,
compared to a predicted ratio from the single-$\beta$ fit of
\fourstarpredratio$\pm$\fourstarpredratiounc\ with $\sint$
(\nunderpredictsigma$\sigma$), or
\fourstarpredrationosig$\pm$\fourstarpredrationosigunc\ without $\sint$
(\nunderpredictsigmanosig$\sigma$).
With that said, we cannot formally exclude the possibility that
TIC~300651846 is a low-$\beta$ star with no extinction law break.

\subsection{No evidence for extreme starspot filling factors}
\label{subsec:tspec_results}

The LP~12-502 TripleSpec spectra show no evidence for a
spectroscopically distinct cool spot component.  Fits with and without
the $T_{\rm spot}$ term in Equation~\ref{eq:tspec} yield
indistinguishable $\chi^2_\nu$, and in the two-temperature fits the
posteriors collapse to $T_{\rm spot} \approx T_{\rm phot}$.
Fully-convective $\approx$3100\,K stars at this age have
spot-to-photosphere temperature ratios of $\approx$0.9 and filling
factors $f_{\rm spot}$ of 0.1--0.6 \citep{PerezPaolino2025}.  Spots
with lower contrast are harder to detect through this method.  Since these models recover
filling factors down to $f_{\rm spot}$$\approx$0.2
\citep{PerezPaolino2025}, we adopt $f_{\rm spot} \lesssim 0.2$ as the
limit for spots with temperature contrasts large enough to be
detected on LP~12-502.

The TripleSpec data do, however, require an excess component.
Relative to a photosphere-only model, adding it lowers $\chi^2_\nu$
from 1.38 to 0.90, and is favored at $\Delta{\rm BIC}$=1372 and
$\Delta \ln Z$=686.  The component has $T_{\rm excess} = 3827 \pm
271$\,K and $f_{\rm excess} = 0.17 \pm 0.08$ (median and scatter
across spectra), and has an emitting area equivalent to a
0.13\,$R_\odot$ sphere.  With it, the photosphere has $T_{\rm phot} =
3074 \pm 50$\,K, consistent with Table~\ref{tab:starparams}, and
$R_\star$$\approx$0.32\,$R_\odot$, slightly below the SED-based value;
without it, $T_{\rm phot}$ (based on TripleSpec alone) rises to
$\approx$3200\,K and the fit degrades.  The component is hotter than
the photosphere and survives when a cool-spot term is allowed, so it
is not starspot contamination; it seems too hot to be thermal dust
emission; its persistence over four hours argues against a flare; and
its emitting area is far too small for a stellar companion of this
temperature.  None of the $\approx$100 stars fit with the same models
\citep{PerezPaolino2025,PerezPaolino2025b,PerezPaolino2025c} show a comparable excess.
The excess may indicate warm, optically thick material above the
photosphere, though we cannot exclude systematic errors in the
synthetic atmospheres.  Distinguishing these possibilities, and
establishing whether the excess is generic to CPVs, would require
near-infrared spectroscopy of additional CPVs with simultaneous
coverage blueward of 1\,$\mu$m, where the component's fractional
contribution to the flux is largest.

\section{Discussion}
\label{sec:disc}

\begin{figure}[!t]
  \centering
  \includegraphics[width=0.44\textwidth]{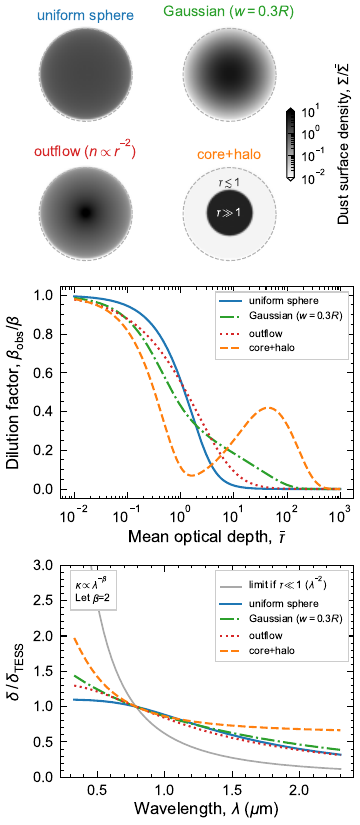}
  \vspace{-0.2cm}
  \caption{{\bf Partial opacity dilutes chromaticity.}
  {\it Top}: surface density maps for four clump
  profiles.  The ``uniform sphere'' has constant number density; the
  ``core+halo'' (Equation~\ref{eq:corehalo}) has $A_{\rm h}/A_{\rm
  c}$=3.
  {\it Middle}: the factor by which an intrinsic dust opacity
  $\kappa\propto \lambda^{-\beta}$ is diluted into an observed dip
  chromaticity, $\delta \propto \lambda^{-\beta_{\rm obs}}$
  (Equation~\ref{eq:betaobs}).  The mean optical depth, $\bar{\tau}$,
  is averaged over the clump's projected disk.
  At fixed $R$ and $\kappa$, profiles with equal
  $\bar{\tau}$ contain equal dust mass.
  {\it Bottom}: example spectra for $\beta$=2; the
  smooth profiles set $\bar{\tau}_{\rm TESS}$=3, and the core+halo
  profile sets $\tau_{\rm h}$=0.1 at 1.2\,$\mu$m.  Only the profile
  that is bimodal
  in surface density matches Figure~\ref{fig:beta}.
  }
  \label{fig:betadilution}
\end{figure}

\subsection{An opaque core and an optically thin halo of dusty plasma}
\label{sec:dustyplasma}

It is tempting to read Table~\ref{tab:fitresults} and
Figure~\ref{fig:beta} as measurements of the dust opacity law,
$\kappa(\lambda)$, whose shape is connected to the size distribution
and composition of the dust grains.  While $\delta(\lambda)$ and
$\kappa(\lambda)$ are related, there is a nuance: the mapping is only
direct when the occulting material is optically thin.  Nothing about
the data guarantees that we are in this regime.  Consider an absorbing
clump of material occulting a uniform stellar disk of area $A_\star$.
The dip depth $\delta$ is
\begin{equation}
  \delta(\lambda) = \frac{1}{A_\star}
  \int \left[ 1 - e^{-\kappa \Sigma} \right] {\rm d}A,
  \label{eq:depthintegral}
\end{equation}
where $\Sigma$ is the surface density of the dust, $\kappa$ is the
opacity (and only wavelength-dependent term), and the integral is over
the dust clump's area.  Approximating $\kappa(\lambda) \propto
\lambda^{-\beta}$ in the optical and near-infrared, one can show that
the {\it observed} slope $\beta_{\rm obs}$ from $\delta(\lambda)$ in
Figure~\ref{fig:beta} is related to the {\it underlying} dust opacity
law's $\beta$ through
\begin{equation}
  \beta_{\rm obs} \equiv
  -\frac{{\rm d} \ln \delta}{{\rm d} \ln \lambda}
  = \beta \, \frac{{\rm d} \ln \delta}{{\rm d} \ln \kappa}.
  \label{eq:betaobs}
\end{equation}
The derivative on the right-hand side is a ``dilution factor'' which
lies between 0 and 1 for any surface density distribution.  The
observed dip-depth $\beta_{\rm obs}$ of Table~\ref{tab:fitresults} and
Section~\ref{subsec:beta} therefore provides only a lower bound on the
dust opacity's $\beta$, with the two related by
Equation~\ref{eq:betaobs}. 

% NOTE:
% each Σ scale in the distribution produces one bump centered at τ ~ 1
% for that scale. the bimodal surface density distribution, with
% well-separated scales, gives two bumps with a valley between.

Figure~\ref{fig:betadilution} quantifies this damping for
representative surface-density profiles and mean optical depths.  Note
that while the ``uniform sphere'' profile has $\Sigma(r) \propto (R^2
-r^2)^{1/2}$, this is not visible in the top panels due to the color
stretch.  The Gaussian has width $w$ such that $\Sigma(r) \propto
\exp(-r^2 / (2 w^2))$.  The middle panel shows that for the smooth
profiles, by a mean optical depth $\bar{\tau}$$\approx$1 the observed
slope falls to 40--60\% of the intrinsic slope, and by
$\bar{\tau}$$\approx$3 to 20--30\%.  The lower panel gives example
vertical slices from the middle, setting $\bar{\tau}$$=$3 for the
smooth profiles and exploring the impact of dilution on a ``steep''
$\beta$=2 intrinsic opacity.  All models flatten the extinction
curve.  However, even after accounting for this dilution, comparison
with Figure~\ref{fig:beta} shows that the smooth profiles provide a
poor fit to the data: their blue dips are not deep enough, and their
$K$-band dips are too shallow.

If dust provides the opacity, the simplest physical interpretation of
a broken power-law in $\delta(\lambda)$ is therefore to invoke a {\it
bimodal} surface density distribution: an opaque core of projected
area $A_c$ ($\tau \gg 1$ out through $K$ band) surrounded by an
optically thin halo of area $A_h$ and depth $\tau_h(\lambda)$.  Such a
distribution yields
\begin{equation}
  \delta(\lambda) \approx \frac{A_c}{A_\star}
  + \frac{A_h}{A_\star} \left[ 1 - e^{-\tau_h(\lambda)} \right],
  \label{eq:corehalo}
\end{equation}
such that the infrared depths sit on the core's gray floor while the
optical depths are dominated by the halo's chromatic term, with the
break falling near where the two terms are comparable.  The
``core+halo'' curve in the lower panel of
Figure~\ref{fig:betadilution} shows an example with $A_{\rm h}/A_{\rm
c}$=3 and $\tau_{\rm h}(1.2\,\mu{\rm m})$=0.1: the same $\beta$=2
opacity law that the smooth profiles flatten into a gray curve instead
yields a broken power law resembling the data.  A brute-force
exploration of the parameter space indicates an allowed parameter
range of $1 \lesssim A_{\rm h}/A_{\rm c} \lesssim 30$,
$10^{-4}\lesssim \tau_{\rm h}\lesssim 0.3$ at 1.2\,$\mu$m, and
$1<\beta<4$, where the upper bound on $\beta$ is the
Rayleigh-scattering limit.  This range of $\beta$ excludes 
a {\it halo} dust population dominated by mm to cm-sized solids, but is
otherwise consistent with a wide range of dust compositions and size
distributions.
The {\it core} material is required by Equation~\ref{eq:corehalo} only to be
opaque, not to be contiguous or centrally concentrated: a swarm of
macroscopic solids dispersed through the halo would yield the same
$\delta(\lambda)$.   However, a pebble swarm configuration seems
implausible because magnetic coupling fails above grain
sizes of order 10\,$\mu$m \citep[see Equation~19 of][]{Sanderson2023};
we therefore favor a ``core'' composed of the same small grains as the
halo, but at high column density (either centrally concentrated, or
with multiple high-$\Sigma$ patches), rather than a distinct population of
large solids.

\subsection{Where do the dust and plasma come from?}
\label{sec:origin}

\begin{figure}[!t]
  \centering
  \includegraphics[width=0.48\textwidth]{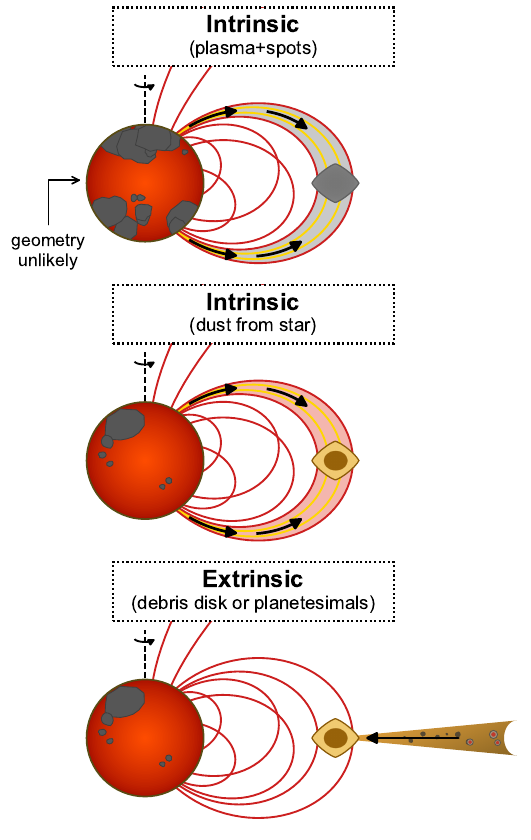}
  \vspace{-0.4cm}
  \caption{{\bf Dust is required in the viable origin scenarios.}
  {\it Top}: corotating plasma clumps plus high starspot filling
  factor and no dust.
  {\it Center}: Dust condensing from magnetically confined stellar
  plasma.
  {\it Bottom}: Dust captured from an extrinsic reservoir (debris
  disk or undetected planetesimals).
  The dust-free scenario is disfavored; the lower two remain
  possible, and differ in where the dust originates.  At least some
  starspots are present in all scenarios.  The stars must have
  ``centrifugal'' magnetospheres with magnetospheric radii
  $R_{\rm m}$ exceeding their corotation radii $R_{\rm c}$; the
  clumps remain long-lived only between the two extremes.
  }
  \label{fig:cartoons}
\end{figure}

Where does the occulting material responsible for CPV dips come from?
Figure~\ref{fig:cartoons} sketches three possible scenarios: {\it
(i)} dust-free plasma clumps combined with starspots, {\it (ii)} dust
that condenses out of the magnetically confined stellar plasma, and
{\it (iii)} dust captured from an extrinsic reservoir, such as a
debris disk or undetected planetesimals.  All three scenarios require
material in corotation beyond $R_{\rm c}$, held in the local
potential minima along closed field lines \citep{Townsend2005}, with
$R_{\rm m}$$>$$R_{\rm c}$ so that magnetic tension enforces
corotation at the clump locations.  The sinusoidal Balmer emission
supports this: for TIC~300651846 and TIC~141146667, the clumps emit
at $\approx$1.1--2.1\,$R_{\rm c}$.  Starspots are present in every scenario, since
CPVs are young, magnetically active, and show smooth rotational
modulation over which the dip signals are superposed.  While dark
facular networks could also be present \citep{Shapiro2026}, they are
relevant only in that they could influence this smooth variability.

In scenario {\it (i)}, the plasma clumps condense out from the star
\citep[e.g.][]{Daley-Yates2024}, and a very specific starspot geometry
is required.  The chromaticity must be provided by the plasma, the
spots, or a combination thereof.  The plasma cannot do it alone:
neither H$^-$ opacity nor hydrogen bound-free opacity provides the
correct functional form.  The most likely plasma opacity source would
instead be Thomson scattering from the free electrons, combined with
the ``transit light source effect'' \citep{Rackham2018}, in which
unocculted spots outside the transit chord provide the apparent
chromaticity.  Exploring such models, however, shows that the required
spot-covering fractions are 70--90\% of the stellar area, while the
clump sizes required by CPV dip depths ($R/R_\star$$\gtrsim$0.1) leave
little room for clear transit chords.  The geometric probability of
clear chords across the entire ensemble of observed chromatic CPV dips
is low enough to rule this scenario out.  Separate arguments against
it include the lack of light curve response to changes in
H$\alpha$ emission (Section~\ref{subsec:mage_interpret}), and the
evidence from TripleSpec for starspot covering fractions
$\lesssim$20\% in LP~12-502 (Section~\ref{subsec:tspec_results}).

While we disfavor scenario {\it (i)}, the data remain compatible with
scenarios {\it (ii)} and {\it (iii)}.  In either, the dust must become
collisionally charged \citep{Sanderson2023} to remain magnetically
confined \citep{Daley-Yates2026}.  The two scenarios may differ in
their expected extent of mixing between the gas and dust, and in the
composition of the grains.  Either origin channel would yield new
information: in scenario~{\it (ii)} we would learn that young M dwarfs
condense out dust, similar perhaps to how this process occurs in the
atmospheres of AGB stars \citep{Hofner2018}; in scenario~{\it (iii)}
we would gain a new way of identifying and studying the detritus of
planet formation.

\subsection{Continued relevance of PTFO\,8-8695}
\label{sec:ptfo}

Of the stars in Figure~\ref{fig:beta}, PTFO\,8-8695 (CVSO~30) has
the longest and most contested observational record.
To recap: \citet{vanEyken2012} identified a hot
Jupiter candidate in Orion; \citet{Barnes2013} and \citet{Kamiaka2015}
subsequently argued that its orbit was nodally precessing, and
\citet{Johns-Krull2016} found that the source showed
sinusoidal-in-time H$\alpha$ emission at circumstellar velocities.
Photometric follow-up found that the system's period
was fixed, but that the signal jumped around in orbital phase and
split into multiple dips per cycle
\citep{Ciardi2015,Yu2015,Raetz2016,Onitsuka2017,Tanimoto2020}.
\citet{Bouma2020} used TESS and Gaia to argue that the system shows
multiple indicators for being an intermediate-separation (multiple~AU)
binary, one component of which shows an ordinary rotation signal, and
the other component being a complex periodic variable.  While this
interpretation may not be uniformly accepted \citep{Grosson2021},
further photometric follow-up by \citet{Bischoff2022} has been broadly
consistent.

The characterization of PTFO 8-8695 provided photometric and
spectroscopic details that have been remarkably prescient for
understanding CPVs.  Spectroscopically, \citet{Johns-Krull2016} showed
evidence for Balmer variability similar to that seen for TIC~141146667
and TIC~300651846, and consistent with the idea of bound circumstellar
plasma being present.  Photometrically, \citet{Tanimoto2020} monitored
the source for five years in optical and infrared bands, and observed
a break in the chromaticity of this system's dip depths going out to
$K$-band similar to that reported in this work.  One of their proposed
resolutions was the scenario of an opaque dust core with an optically
thin dust halo; this is the same scenario that we favor in
Section~\ref{sec:dustyplasma}.  The Spitzer 4.5\,$\mu$m
light curve observed by \citet{Ciardi2015} showing a 0.5\% deep dip in
2012 remains a unique dataset, as it provides the only information we
have about the chromaticities of any CPV dips redward of $K$-band.

\subsection{The chromaticity differs from analog magnetic rotators}
\label{sec:analogs}

Massive magnetic B stars are the only class of star known to exhibit
similar light curves to CPVs.  The prototype, $\sigma$ Ori~E, has a
long history \citep[e.g.][]{Landstreet1978,Townsend2005b,Oksala2015},
but the summary is that it hosts a centrifugally supported
magnetosphere in which a long-lived ``pinched'' plasma torus
\citep{Townsend2008} produces photometric eclipses and
sinusoidal-in-time spectroscopic emission.  \citet{Bouma2024} argued
that the additional step to produce ``complex'' light curves comes
from magnetic topology, on the basis that the few $\sigma$ Ori~E
analogs showing complex light curves also show (rare) evidence from
Zeeman Doppler imaging for non-dipolar magnetic fields.  The analogy
extends to the radio: $\sigma$ Ori~E shows rotationally modulated
gyrosynchrotron emission from its corotating magnetosphere
\citep{Leto2012}, and polarized, variable radio emission has now been
reported for two CPVs, DG~CVn \citep{Kaur2025} and TIC~146539195
\citep{Kaur2024,Curiel2026}.

The physical analogy breaks down in that the chromaticities of the
$\sigma$ Ori~E eclipses are different from what is seen for
CPVs.  \citet{Hesser1977} and \citet{Oksala2007} provided photometry
that showed near-gray chromaticity in $vby$, with a slight depth
decrease in $v$ relative to $y$, and a major depth increase in $u$.
This behavior mirrors the hydrogen bound-free opacity curve as a
function of wavelength \citep{Gray2008}: the Balmer break at
3646\,\AA\ captures the $u$ band, while $vby$ would have a slight
decrease before the break.  This is consistent with the opacity in the
$\sigma$ Ori~E stars coming purely from hydrogen.  While
$u$-band photometry of CPVs does not exist, their chromaticity across
$griz$ is inconsistent with the dip depths predicted by
bound-free opacity.

\section{Conclusions}
\label{sec:conc}

CPVs show smooth rotational modulation together with sharp dips from
circumstellar material.  The material must be circumstellar given how
quickly the flux changes, and magnetically bound, given its
synchronicity with the stellar rotation period and its response to
flares \citep{Stauffer2017,Bouma2024}.   A priori, the material could
be gas ({\it i.e.}, plasma), dust, or both.  However, in
Section~\ref{sec:origin} we rule out the ``gas-only'' scenario:
neither H$^-$ nor hydrogen bound-free opacity yields the observed
wavelength dependence, and the remaining option, Thomson scattering
with chromaticity supplied by unocculted starspots, requires spot
covering fractions that are excluded.  While gas is conspicuously
present, dust must provide the opacity.

Gas is conspicuous for two reasons: sinusoidal-in-time Balmer emission
out of transit, and Balmer dimming during transit.  The emission
yields a measurement of the hydrogen number density of the bound
material at a few stellar radii.  After TIC~141146667 ($n_{\rm
H}$$\approx$1$\times$10$^{11}$\,cm$^{-3}$; \citealt{Bouma2025}),
TIC~300651846 shows emitting plasma at $n_{\rm
H}$$\approx$3--8$\times$10$^{11}$\,cm$^{-3}$
(Section~\ref{subsec:mage_interpret}), and LP~12-502, observed at
comparable sensitivity, gives $n_{\rm
H}$\,$\lesssim$\,5$\times$10$^{10}$\,cm$^{-3}$
(Appendix~\ref{app:keck_ootspec}).  The range for TIC~300651846
reflects a six-fold change in clump H$\alpha$ luminosity, during which
the simultaneous TESS light curve did not discernibly change.  The dip
morphology is therefore decoupled from the plasma density, pointing to
dust as the main opacity source.  Balmer dimming during broadband
dips, meanwhile, appears in every CPV observed spectroscopically
during a dip, with repeatable timing offsets of tens of minutes in two
stars (Section~\ref{subsec:ha_flux_dip_interpret}).  The gas and dust
are spatially associated, but the dust column does not track the gas
density.

Broadband chromaticity currently provides the best information about
the dust.  We combined our new observations with the literature and
fitted four variants of a $\delta \propto \lambda^{-\beta}$ extinction
curve.  While all four models agree that the dips are chromatic, a
simple power law provides a poor fit ($\chired$$\gg$1).  Two
modifications can improve it: the extinction law can flatten in the
near-infrared, or there can be star-to-star or epoch-to-epoch
variability in $\beta$.  The data currently cannot separate the two
due to the limited resolution of multi-band photometry: future
{\it spectrophotometry} would solve this issue.  If at present we
assume a universal law, then a broken power-law is preferred, with
$\beta_1 = \brokenbetaone \pm \brokenbetaoneunc$ blueward of
$\lambda_b = \brokenlambdabreak \pm \brokenlambdabreakunc$\,$\mu$m and
$\beta_2 = \brokenbetatwo \pm \brokenbetatwounc$ redward.

The break, rather than the slopes, carries most of the physical
information (Sections~\ref{sec:dustyplasma} and~\ref{sec:origin}).
It can be produced by a bimodal distribution of dust surface
densities: an opaque ``core'' 
and a nearly transparent ``halo''.
In this picture, the infrared depths sit on the core's gray floor
while the optical depths are set by the halo's chromatic term.
We find that such a model reproduces the data
for halo to core area ratios $1 \lesssim A_{\rm h}/A_{\rm c} \lesssim
30$ and halo optical depths $10^{-4}\lesssim \tau_{\rm h}\lesssim 0.3$
at 1.2\,$\mu$m.
The {\it observed} optical chromaticity $\beta_1$ then provides a lower
bound on the {\it intrinsic} spectral index of the halo dust opacity: for
$\kappa \propto \lambda^{-\beta}$ we infer $1<\beta<4$.  This excludes
a halo dominated by mm to cm-sized solids, but is otherwise
consistent with a wide range of dust compositions and size
distributions.  
The core's grain size is not constrained by the chromaticity, but
macroscopic solids could not remain magnetically coupled, so the core
most plausibly consists of the same small grains at high column
density.

Gas is conspicuous: it emits in corotation, dims during transit, and
changes in density without altering the dips.  Dust is cryptic: it
must supply the opacity, yet whether it condenses from the star or is
captured from planetesimal debris (Section~\ref{sec:origin}) is
unknown.  The core+halo picture of a dusty plasma makes three key
predictions.  First, dips should be gray redward of $\lambda_b$ out to
whatever wavelength the core becomes optically thin.  Second, dip
depths should vary smoothly across the Balmer break, unlike the
$u$-band jump of the hydrogen-dominated $\sigma$ Ori~E analogs
(Section~\ref{sec:analogs}).  And third, if the grains contain
silicates, the 10\,$\mu$m Si-O stretching feature should appear as
excess depth above the gray floor, with a profile that could
distinguish wind condensates from collisional debris.

\begin{acknowledgements}
  \newline
  L.G.B. acknowledges support from the Carnegie Fellowship, the
  Heising-Simons 51 Pegasi~b Fellowship, and NASA Grant 80NSSC24K0474
  during this work's data collection.
	J.G.M. acknowledges support from the Heising-Simons Foundation and
	the Pappalardo family through the MIT Pappalardo Fellowship in Physics.
  We thank the staff of Magellan, Keck, and Palomar
  Observatories, including the telescope operators and instrument
  specialists, for their support of the observations presented in this
  work; we also thank Eric Persson and Andy Monson for assistance with FourStar
  planning.
  This work used data from TESS \citep{TESS2min}.
  % via https://heasarc.gsfc.nasa.gov/docs/tess/approved-programs.html
  The 2-minute cadence observations used in this work were proposed by
  the guest observer (G) and director's discretionary time (DDT)
  programs listed below; we used light curves from SPOC
  \citep{Jenkins2016}, TARS \citep{Boyle2026}, and unpopular
  \citep{Hattori2022}.
  % 300651846 S98
  {\it TIC~300651846:}
  % 300651846 S87 + S88 G07140;G07036;G07103
  G07140 (A.~Boyle), G07036 (T.~Wainer), G07103 (L.~Bouma);
  G08046 (M.~Battley), G08137 (A.~Feinstein).
  % DG CVn S50
  {\it DG~CVn:}
  % G04212;G04132;G04211;G04188;G04039;G04191;20SECONDTARGET
  G04212 (R.~Paudel), G04132 (W.~Howard), G04211 (F.~Marocco), G04188
  (M.~Silverstein), G04039 (J.~Davenport), G04191 (J.~Burt).
  % LP 12-502 / 402980664 s58
  {\it LP~12-502:}
  G05106 (E.~Gillen), G05152 (R.~Cloutier), 
  % LP 12-502 s73
  G06030 (L.~Bouma), G06120 (E.~Gillen),
  % LP 12-502 s86
  G07089 (E.~Gillen), G07103 (L.~Bouma).
  {\it TIC 262400835:}
  DDT028 (L.~Bouma).
  Funding for the TESS mission is provided by NASA's Science Mission
  directorate.
  The HIRES data were obtained at the Keck Observatory.
  We recognize the importance that the summit of Maunakea has always
  had within the indigenous Hawaiian community, and we are grateful
  for the opportunity to conduct observations from this mountain.
	This work makes use of observations from the Tierras Observatory, which is
	supported by the National Science Foundation under Award No. AST-2308043.
	Tierras is located within the Fred Lawrence Whipple Observatory; we thank all
	the staff there who help maintain this facility.
  This article is also based on observations made with the MuSCAT2
  instrument, developed by ABC, at Telescopio Carlos Sánchez operated
  on the island of Tenerife by the IAC in the Spanish Observatorio del
  Teide.  This work is partly supported by JSPS KAKENHI Grant Numbers
  JP24H00017, JP25K24620, JP26H01402, JP24K00689, JP25K17450,
  JP26K00755 and JP26K17193.
  We also acknowledge financial support from the Agencia Estatal de
  Investigaci\'on of the Ministerio de Ciencia e Innovaci\'on
  MCIN/AEI/10.13039/501100011033 and the ERDF ``A way of making
  Europe''
  through projects PID2021-125627OB-C32 and PID2024-158486OB-C32. This
  work is supported by the ERC Grant (ERC Advanced Grant SPEAR, GA
  101200674), funded by the European Union.  Views and opinions
  expressed are however those of the author(s) only and do not
  necessarily reflect those of the European Union or the  European
  Research Council Executive Agency (ERCEA). Neither the European
  Union nor the granting authority can be held responsible for them.
\end{acknowledgements}

{\it \large Contributions}: 
LGB led the design, execution, and interpretation of the
project.  FPP performed the TSPEC analysis.  Data were acquired and reduced by
JHL, KI, NN, and AF (MuSCAT1); EP, FM (MuSCAT2); JGM, PT, DC (Tierras); AB
(KeplerCam); FPP (TSPEC); and LGB (MIKE, MagE, FourStar, HIRES, DBSP, TSPEC).
Observations by JJS (FourStar) and ASC (HIRES), and reduction assistance by HI
(HIRES) enabled the work.  Finally, FPP, LAH, JJS, MMJ, DRC, and JAV
contributed to the interpretation of the results, and all authors
reviewed the manuscript.

\facilities{
  Magellan:Baade,
  Magellan:Clay,
  Keck:I,
  Hale,
  NAOJ:1.88m,
  Sanchez,
  FLWO:Tierras,
  FLWO:1.2m,
  Gaia,
  TESS,
	2MASS,
	SDSS.
}

\software{
  astroariadne \citep{Vines2022},
  astropy (\citealt{astropy:2013,astropy:2018,astropy:2022}),
  carpy \citep{Kelson2000,Kelson2003},
  Claude \citep{claude2026},
	DBSP\_DRP \citep{dbsp_drp:joss,dbsp_drp:zenodo},
	dustmaps \citep{2018JOSS....3..695M},
	dynesty \citep{Speagle2020},
  matplotlib \citep{matplotlib},
  numpy \citep{numpy},
  PyMC \citep{pymc2023},
	PypeIt \citep{Prochaska2020},
  scipy \citep{scipy}.
}

%%%%%%%%%%%%%%%%%%%%%%%%%%%%%%%%%%%%%%%%%%%%%%%%%%%%%%%%%%%%%%%%%%%%%%%%%%%%%%%
%%%%%%%%%%%%%%%%%%%%%%%%%%%%%%%%%%%%%%%%%%%%%%%%%%%%%%%%%%%%%%%%%%%%%%%%%%%%%%%

\appendix
\section{Spectroscopy of CPVs Outside the Narrow Flux Dips}
\label{app:ootspec}

In the main text we focus on photometric and spectroscopic signatures
of the circumstellar material present around CPVs.  However, due to
the long-term unpredictability in CPV light curve morphology, some
nights yielded coverage of only spot-induced or flare-induced
broadband flux variability without providing any
evidence---spectroscopic or photometric---for bound circumstellar
material.  We describe the results of those observations below.

\subsection{Keck/HIRES}
\label{app:keck_ootspec}

\begin{figure*}[!t]
  \centering
  \includegraphics[width=0.8\textwidth]{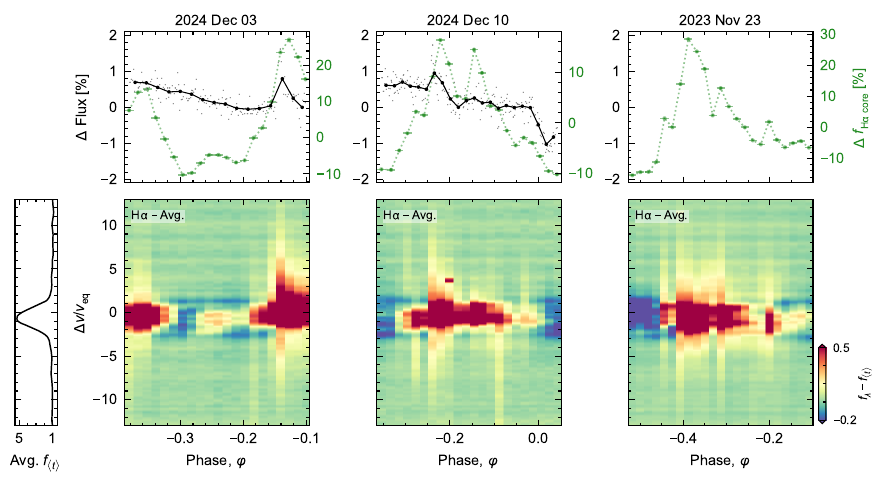}
  \vspace{-0.2cm}
  \caption{{\bf Keck/HIRES H$\alpha$ time series of LP~12-502
  vs.~simultaneous TESS photometry.}  The lower left panel shows the
  time-average H$\alpha$ profile on 3~December 2024.  The 
  bottom columns show the corresponding mean-subtracted H$\alpha$
  profiles.  On top, simultaneous TESS flux measurements are in black,
  with the median-normalized H$\alpha$ line flux at
  $|v/v_{\rm eq}|<1$ in green.
  The variability on all three nights is flare-induced.  The lack of
  (bound) circumstellar emission yields an upper limit on the number
  density of any plasma clumps, discussed in the text.}
  \label{fig:hires}
\end{figure*}

Figure~\ref{fig:hires} summarizes TESS and HIRES data for LP~12-502 on
the nights of 23~November 2023, 3~December 2024, and 10~December 2024.
The static spectra exhibit Balmer series emission up to
$n$=15$\rightarrow$2, strong chromospheric emission in the
\ion{Ca}{2}~H\&K lines, and the expected forest of molecular
absorption lines for a non-accreting (``weak-lined'') pre-main
sequence M4.5Ve dwarf.

The HIRES spectra show flare-induced variability over all three
nights.  The most direct indicator for this variability is in
chromospheric emission lines; excess emission appears during the
variable epochs while the redder orders, including the
order-normalized continuum, appear stable.  This spectral variability
is temporally coincident with elevated TESS flux levels
(Figure~\ref{fig:hires}).  Some of these flares are symmetric in the
blue and red wings of the Balmer lines, consistent with electric
field-induced pressure broadening \citep{Kowalski2017}.  A few however
are asymmetric, suggestive of flare-associated ejection or infall of
cool gas \citep{Vida2016,Vida2019,Maehara2021,Namekata2021,Inoue2023}.
The velocity profiles in H$\alpha$ are mirrored across numerous
chromospheric lines, including the entire Balmer series, the
\ion{Ca}{2}~H\&K lines, \ion{Ca}{1}\,4226, \ion{He}{1}\,4026,
\ion{He}{1}\,4471, the Na doublet, and the Mg triplet.  The flaring
epochs also coincide with the excitation of at least a dozen
\ion{Fe}{1} lines spanning 3820\,\AA--4050\,\AA \citep{NIST_ASD}.
These lines are not present in the quiescent spectra; they appear in
emission with $\Delta\lambda$$\approx$0.2\,\AA\ over the few-hour
duration for which the other flare diagnostics are also present.  This
is consistent with established phenomenology for M dwarf flares
\citep{Fuhrmeister2008}.

% re: iron lines, cf
% https://docs.google.com/spreadsheets/d/18ADpjTs8v_zp22vRA5i7o6JrJvtaq8PYVZeiCXbQo-k/edit?usp=sharing
%
% full details in 20240708_j531_LP12-502_metal_emission_line_analysis.txt

While the HIRES spectra and photometry of LP~12-502 presented abundant
evidence for flare-induced variability, no variability could be
clearly associated with any bound circumstellar hydrogen.
Nonetheless, the same calculation as in
Section~\ref{subsec:mage_interpret} enables a conversion of this 
non-detection of H$\alpha$ emission  to an upper limit on the number
density of bound circumstellar plasma.
% via /Users/luke/Dropbox/proj/hotcold/drivers/calc_number_density_constraint_from_ha_emission.py
For LP~12-502, $L_\star$=0.011$\pm$0.001\,L$_\odot$,
$R_\star$=0.37$\pm$0.01\,R$_\odot$ \citep{Bouma2024}, and we measure a mean
stellar H$\alpha$ EW and scatter of $0.22 \pm 0.02$\,\AA.  The implied
stellar H$\alpha$ line luminosity is
$\approx$3.0$\times$10$^{26}$\,erg\,s$^{-1}$, two orders of magnitude
lower than that of TIC~300651846.
We derive an upper limit on the equivalent width of any excess
H$\alpha$ by injecting synthetic emission into the spectra shown in
Figure~\ref{fig:hires}.
This process indicated that we would have been sensitive to 
circumstellar H$\alpha$ emission with EW $\gtrsim$ 10 m\AA.
As a result, we can rule out plasma clumps with number density $n_{\rm
H} \gtrsim 5 \times 10^{10}$\,cm$^{-3}$, or assuming
$r$$\approx$0.1\,$R_\star$, any clumps with $M_{\rm
gas}$$\gtrsim$6$\times$10$^{15}$\,g.
This limit on the clump hydrogen number density is an order of
magnitude lower than the densities observed for TIC~141146667
($\approx$1$\times$10$^{11}$\,cm$^{-3}$; \citealt{Bouma2025}) and
TIC~300651846 (3--8$\times$10$^{11}$\,cm$^{-3}$;
Section~\ref{subsec:mage_interpret}).

\subsection{Hale/DBSP}
\label{app:dbsp_ootspec}

\begin{figure*}[!t]
  \centering
  \includegraphics[width=0.75\textwidth]{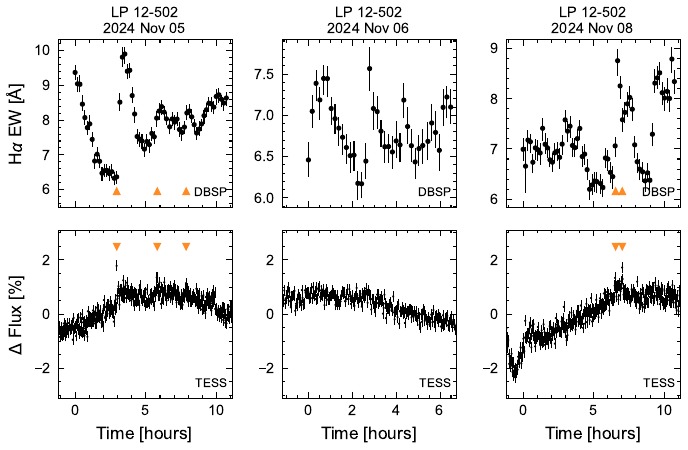}
  \vspace{-0.2cm}
  \caption{{\bf Hale/DBSP H$\alpha$ equivalent widths of LP~12-502
  vs.~simultaneous TESS photometry outside complex dips.}  Panels are
  as in Figure~\ref{fig:dbsp}, but for the three nights with only
  flare- and spot-induced variability.  On Nov~5, flare-induced local
  maxima occur at $t$=3.5, 6, and 8~hours; on Nov~6 a similar maximum
  occurs at $t$=2.6\,hours; and on Nov~8, at $t$$\approx$6.5\,hours.
  The Nov~8 DBSP observations began just after the end of a sharp
  ``complex'' flux dip.  }
  \label{fig:dbsp_supp}
\end{figure*}

\begin{figure*}[!t]
  \centering
  \includegraphics[width=0.48\textwidth]{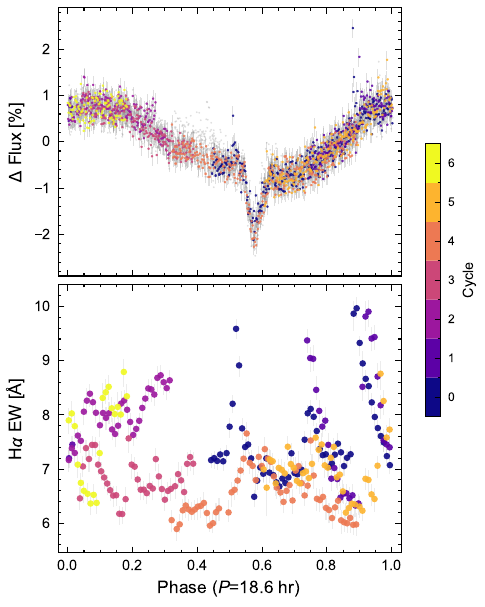}
  \vspace{-0.2cm}
  \caption{{\bf Hale/DBSP H$\alpha$ equivalent widths of LP~12-502
  vs.~simultaneous TESS photometry, 4--8 November 2024.}  Points are
  colored by their orbital cycle number relative to the first
  18.56\,hour period.  Simultaneous TESS measurements use the main
  colormap; TESS data in the 4~November to 8~November baseline that
  were not covered {\it simultaneously} by DBSP due to the day-night
  cycle are shown in gray.
  }
  \label{fig:dbsp_Nov2024}
\end{figure*}

Figure~\ref{fig:dbsp_supp} shows three near-adjacent nights observing
LP~12-502 with Hale/DBSP.  Figure~\ref{fig:dbsp_Nov2024}
gives a complete visualization of the TESS flux and DBSP H$\alpha$~EW
behavior over the five adjacent nights in November~2024, folded on the
18.56\,hour period.  While LP~12-502 exhibited extreme variability
from 2019--2023, showing four to eight sharp flux dips per cycle, it
displayed only one or two sharp flux dips from May to December~2024
(TESS Sectors 78--86).  A movie showing the star's TESS light curve
from 2019--2025 is
\href{https://lgbouma.com/movies/LP_12-502_20250218.mp4}{{\bf
available online}}.  In our November~2024 DBSP run, flare-induced
variability played a significant role in modulating the
H$\alpha$ equivalent width measurements, and only two of the five
nights (discussed in the main text) overlapped with the complex flux
dips that define the CPV class.

\section{Time-Series Photometry Fitting Diagnostics}
\label{app:fit_diagnostics}

\begin{figure*}[!tp]
  \centering
  \includegraphics[width=0.8\textwidth]{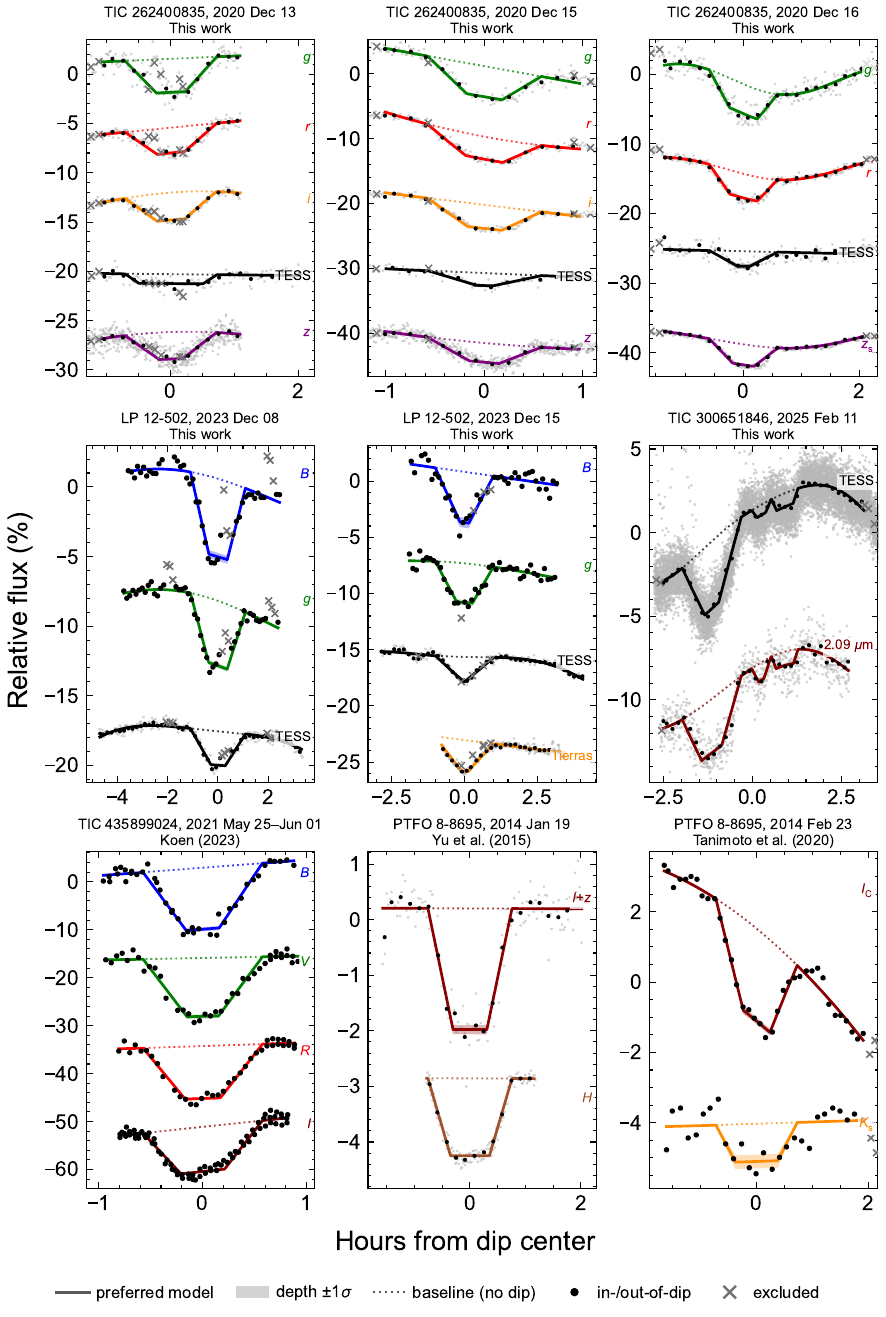}
  \vspace{-0.2cm}
  \caption{{\bf Trapezoidal fits to multi-band CPV photometry (1/2).}
  Data from TIC\,262400835, LP\,12-502, and TIC\,300651846 are new to
  this work.  We digitally transcribed the measurements from
  TIC\,435899024 \citep{Koen2023} and PTFO\,8-8695
  \citep{Yu2015,Tanimoto2020}.  Dip depth uncertainties
  ($\pm$1-$\sigma$) are shown in the underplotted contours.}
  \label{fig:dipfit1}
\end{figure*}

\begin{figure*}[!tp]
  \centering
  \includegraphics[width=0.8\textwidth]{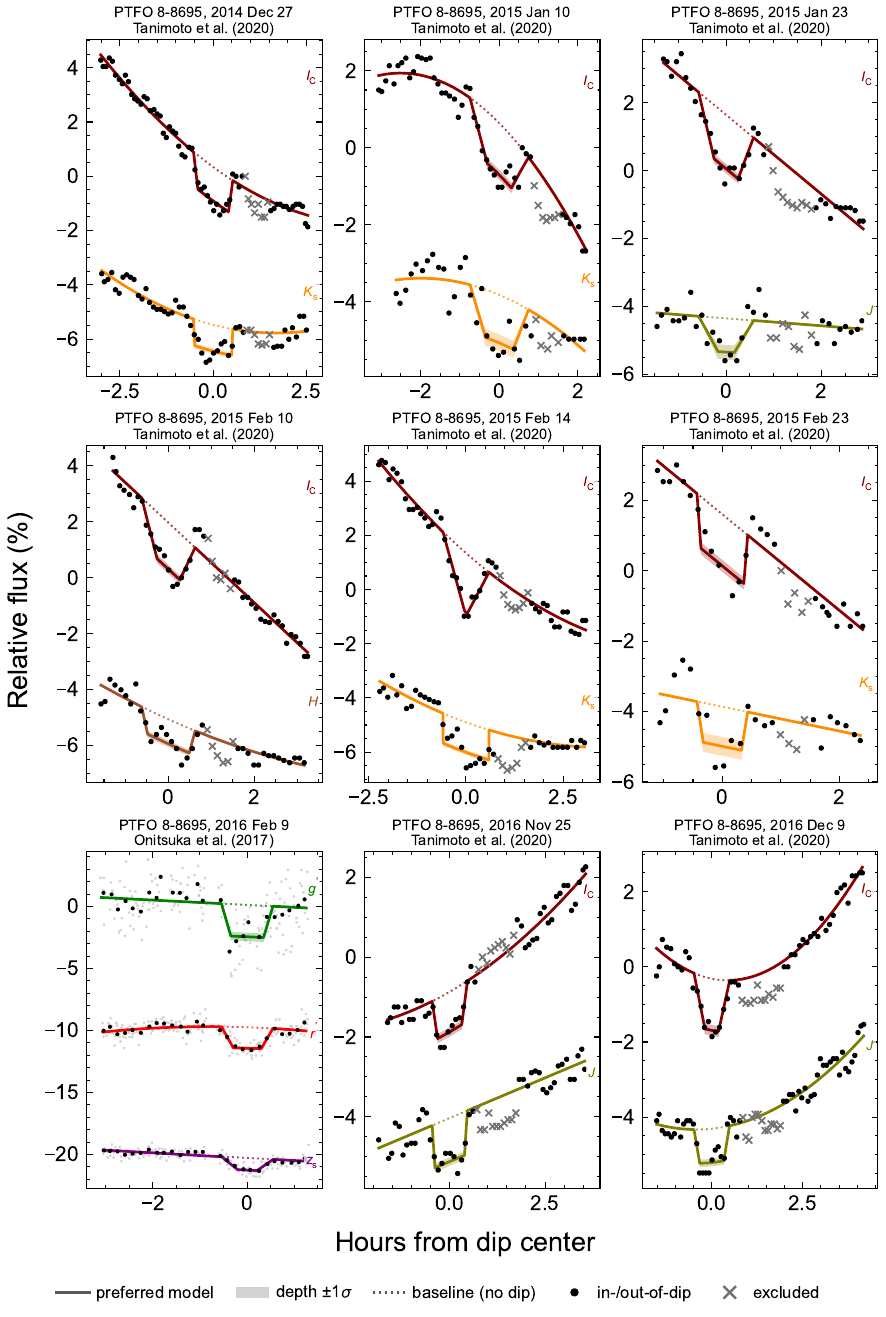}
  \vspace{-0.2cm}
  \caption{{\bf Trapezoidal fits to multi-band CPV photometry (2/2).}
  We digitally transcribed these PTFO\,8-8695 measurements from
  \citet{Onitsuka2017} and \citet{Tanimoto2020}.  The underplotted
  contours show our fitted $\pm$1-$\sigma$ dip depth uncertainties. }
  \label{fig:dipfit2}
\end{figure*}

Figures~\ref{fig:dipfit1} and~\ref{fig:dipfit2} show the results from
our photometry fitting routine, which forms the basis for the
chromaticity analysis in Section~\ref{subsec:beta}.

%%%%%%%%%%%%%%%%%%%%%%%%%%%%%%%%%%%%%%%%%%
%%%%%%%%%%%%%%%%%%%%%%%%%%%%%%%%%%%%%%%%%%

%\clearpage

\bibliographystyle{aasjournalv7.1}
\bibliography{cpvbib.bib}

%%%%%%%%%%%%%%%%%%%%%%%%%%%%%%%%%%%%%%%%%%
%%%%%%%%%%%%%%%%%%%%%%%%%%%%%%%%%%%%%%%%%%

\end{document}